\documentclass{aa}
\usepackage{graphicx}
\usepackage{txfonts}
\usepackage{color}
\usepackage{hyperref}

\newcommand*\setPy{\ensuremath{\mathcal{P}}_\mathcal{Y}}
\newcommand*\setPx{\ensuremath{\mathcal{P}}_\mathcal{X}}
\newcommand*\settildePy{\ensuremath{\widetilde{\mathcal{P}}_\mathcal{Y}}}
\newcommand*\settildePx{\ensuremath{\widetilde{\mathcal{P}}_\mathcal{X}}}

\newcommand*\setTy{\ensuremath{\mathcal{T}}_\mathcal{Y}}
\newcommand*\setTx{\ensuremath{\mathcal{T}}_\mathcal{X}}

\DeclareMathOperator*{\argmin}{arg\,min}

\usepackage{etoolbox}

\begin{document}

    \title{TRAP: A temporal systematics model for improved direct detection of exoplanets at small angular separations}

\author{M. Samland\inst{1, 2, 3}
\and
J. Bouwman\inst{1}
\and
D. W. Hogg\inst{4, 5, 1, 6}
\and
W. Brandner\inst{1}
\and
T. Henning\inst{1}
\and
M. Janson\inst{3}
}

\institute{Max-Planck-Institut f\"ur Astronomie, K\"onigstuhl 17, 69117 Heidelberg, Germany\\
\email{samland@mpia.de}
\and
International Max Planck Research School for Astronomy and Cosmic Physics at the University of Heidelberg (IMPRS-HD)
\and
Department of Astronomy, Stockholm University, Stockholm, Sweden
\and
Center for Cosmology and Particle Physics, Department of Physics, New York University, 726 Broadway, New York, NY 10003, USA
\and
Center for Data Science, New York University, 60 Fifth Ave, New York, NY 10011, USA
\and
Flatiron Institute, Simons Foundation, 162 Fifth Ave, New York, NY 10010, USA
}

\date{Received 13 December 2019; accepted 23 November 2020}

 
  \abstract
   {High-contrast imaging surveys for exoplanet detection have shown that giant planets at large separations are rare. Thus, it is of paramount importance to push towards detections at smaller separations, which is the part of the parameter space containing the greatest number of planets.
   The performance of traditional methods for the post-processing of pupil-stabilized observations decreases at smaller separations due to the larger field-rotation required to displace a source on the detector in addition to the intrinsic difficulty of higher stellar contamination.}
   {Our goal is to develop a method of extracting exoplanet signals, which improves performance at small angular separations.} 
   {A data-driven model of the temporal behavior of the systematics for each pixel can be created using reference pixels at a different positions, on the condition that the underlying causes of the systematics are shared across multiple pixels, which is mostly true for the speckle pattern in high-contrast imaging. In our causal regression model, we simultaneously fit the model of a planet signal ``transiting'' over detector pixels and non-local reference light curves describing the shared temporal trends of the speckle pattern to find the best-fitting temporal model describing the signal.}
   {With our implementation of a spatially non-local, temporal systematics model, called TRAP, we show that it is possible to gain up to a factor of six in contrast at close separations ($<3\lambda / D$), as compared to a model based on spatial correlations between images displaced in time. We show that the temporal sampling has a large impact on the achievable contrast, with better temporal sampling resulting in significantly better contrasts. At short integration times, (4 seconds) for $\beta$ Pic data, we increase the signal-to-noise ratio (S/N) of the planet by a factor of four compared to the spatial systematics model. Finally, we show that the temporal model can be used on unaligned data that has only been dark- and flat-corrected, without the need for further pre-processing.}
   {}


   \maketitle

%

\section{Introduction}
The field of direct observations of extrasolar planets has seen tremendous progress over the last ten years, both in terms of observational capabilities and of high-contrast imaging data analyses. This is particularly true given the recent advent of instruments dedicated to high-contrast observations such as SPHERE \citep[][]{Beuzit2019}, GPI \citep[][]{Macintosh2014}, and CHARIS \citep{Groff2015}, in addition to the development of more sophisticated observational strategies and post-processing algorithms -- which have paved the way to making it a unique technique for the  study of atmospheres and orbital characteristics of exoplanets directly. However, giant planets and substellar companions at large orbital separations ($\gtrapprox 5 - 10$ au) have been shown to be rare by multiple large direct imaging surveys \cite[e.g.,][]{Brandt2014, Vigan2017, Nielsen2019, Vigan2020arXiv}. Pushing our detection capabilities towards smaller inner-working angles is therefore one of our primary goals, aimed at tapping into a parameter space regime in which planets are more abundant \citep{Nielsen2019, Fernandes2019} and one that overlaps with indirect detection techniques, such as astrometric detections with \textit{Gaia} \citep[e.g.,][]{Casertano2008, Perryman2014} and long-term radial velocity trends \citep[e.g.,][]{Crepp2012, Grandjean2019}.
Detecting planets is intrinsically more difficult the smaller the separation gets, because the speckle background is higher and fewer independent spatial elements are available for statistical evaluation of the significance of the detected signal(s). However, this intrinsic problem is further exacerbated because most algorithms used to model the stellar contamination obscuring the planet signal (the systematics) require strict exclusion criteria that determine which data can be used to construct a data-driven model of these systematics while preventing the incorporation of planet signal.

In this work, we present a novel algorithm designed to improve performance at small angular separations compared to conventional algorithms for pupil-tracking observations. Especially as coronagraphs become more powerful and allow us to probe smaller inner working angles (IWA\footnote{The IWA is defined as the angular separation at which the coronagraphic transmission drops below 50\%.}), the algorithmic performance will become more and more important. We achieve this goal by building a data-driven, temporal systematics model based on spatially non-local data. This model replaces the temporal exclusion criterion with a less restrictive spatial exclusion criterion and allows for the use of all frames in the observation sequence regardless of angular separation. This, in turn, allows us to make better use of the temporal sampling of the data and to create a systematics model that is sensitive to variations on all  timescales sampled. We further employ a forward model of the companion signal, fitting both the systematics model and the planet model at the same time to avoid overfitting and biasing the detection.

In Section~\ref{sec:pupilstab}, we give an overview of systematics modeling in high-contrast imaging data, with a focus on the spatial systematics modeling approach that is traditionally used in pupil-tracking observations. In Section~\ref{sec:causal_modeling}, we motivate our non-local, temporal systematics approach. Our causal regression model exploits the fact that pixels share similar underlying causes for their systematic temporal trends. We show how we can apply such a model to pupil-tracking, high-contrast imaging data. In Section~\ref{sec:data}, we discuss the data used to demonstrate the performance of our algorithm TRAP. Section~\ref{sec:paper2_results} shows and compares the results we obtained. We end with a discussion and a future outlook on how the algorithm can be further improved in Section~\ref{sec:discussion} and we provide a summary of our conclusions in Section~\ref{sec:paper2_conclusions}.

\section{Systematics modeling in pupil-tracking observations}
\label{sec:pupilstab}
The main challenge in high-contrast imaging is distinguishing a real astrophysical signal from the light of the central star, the so-called speckle halo. This stellar contamination is generally orders of magnitude brighter than a planetary companion object in raw data and furthermore, it can  appear locally to be the same as a genuine point source. 

In order to separate the astrophysically interesting signal\footnote{For simplicity, we may simply refer to a planet as the astrophysical signal of interest, as this is the focus of this work, but the same applies for more massive objects and other signals, whether that may be an extended object or a point source.}, such as a planet, from the systematic stellar-noise background, we need to model these systematics. Due to the complexity of the systematics, this is usually done using a data-driven approach, that is, by using the data itself as a basis for constructing the model. In order for this to work, there must be one or more distinguishing properties between the planetary signal and the contaminating systematics (the star's light). These distinguishing properties are also referred to as diversities.

For exoplanet imaging, the primary diversity is spatial resolution, that is, a discernible difference in the position of the companion and the host-star signal on the sky. The most commonly used high-contrast imaging strategy enhances this distinguishing spatial property by inducing a further time-dependence of the measured companion signal by allowing the field-of-view (FoV) to rotate over the course of an observation sequence, while the speckle halo remains stationary. This mode of observation is called pupil-tracking mode and is the basis for algorithms such as angular differential imaging \citep[ADI,][]{Marois2006}.

In this paper, we show that instead of building a spatial speckle pattern model for each image, we can build a temporal light-curve model for each pixel, which provides an alternative, novel way of reducing data taken in pupil-tracking mode.\\

Due to the field-of-view rotation, temporal variation is induced in the planetary signal as measured by the detector, thus creating a signal that varies in both space and time. Because there are two ways in which the planet signal differs from the stellar systematics, we have the freedom to build our systematics model on either, the spatial correlation between images or the temporal correlations between pixel time series (light curves). A mixture model of both presents another possibility which is not explored in this work.

The difference between the two pure approaches can be understood on the basis of whether we treat the data as being either: a) a set of image vectors that are to be explained as a combination of other image vectors taken at other times (spatial-systematics model); or b) a set of time-series vectors that are to be explained as a combination of other time-series vectors taken at other locations (temporal-systematics model). 
Before we go into detail about the proposed temporal systematics approach in Section~\ref{sec:causal_modeling}, we first summarize the commonly used spatial-systematics approach and discuss its drawbacks.

\subsection{State-of-the-art approaches}
Following the invention of roll deconvolution (or roll angle subtraction) for space-based observatories \citep{Mueller1985}, classical ADI \citep{Marois2006} for ground-based observatories using pupil-tracking, and its various evolutions, which initiated the employment of optimization and regression models in the form of a locally optimized combination of images \citep[LOCI, e.g.,][]{Lafreniere2007, Pueyo2012, Marois2014, Wahhaj2015}, virtually all papers and algorithms were focused on using the spatial-systematics approach. In this approach, the training data is typically taken from the same image region where the planet signal is located (as implied also by the naming of LOCI) and the spatial similarity between an individual image and the training set of images displaced in time is used to remove the quasi-static speckle pattern which covers the area where the planetary companion signal is located at that time.
This is the case for most commonly used algorithms, regardless of the implementation details of the model construction. In some cases, the regression is not performed on the images themselves, but a representation of the image vectors in a lower dimensional space using dimensionality reduction techniques, such as a principal component analysis \citep[PCA/KLIP,][]{Amara2012, Soummer2012}. Implementations in publicly available pipeline packages follow this trend, for instance, PynPoint \citep[PCA,][]{Amara2012, Stolker2019}, KLIP \citep{Soummer2012, Ruffio2017}, VIP \citep{Gonzales2017}. The ANDROMEDA algorithm is similarly based on a spatial model, using difference images to suppress the stable features in the speckle pattern \citep{Cantalloube2015} together with a maximum-likelihood approach to model the expected companion signal in the difference images. Another recent approach, called patch covariance \citep[PACO,][]{Flasseur2018}, is aimed at statistically distinguishing a spatial patch that is co-moving with a planet signal based on the properties of a stationary patch.

Two notable exceptions to this spatial approach is the wavelet-based temporal de-noising approach \citep{Bonse2018}, which, however, does not attempt to create a causal regression model of the systematics. It applies a temporal filter and pre-conditions the data before applying a spatial systematics approach. Additionally, the STIM-map approach \citep{Pairet2019} adjusts the detection map based on the temporal residuals left over after a spatial model is applied.

\subsection{Spatial systematics models}
Let us assume we have a data cube, $\tens{D,}$ with a time series of image data\footnote{We use a single index to denote the position (pixel) in the 2D image.} $d_{ik} \in \mathbb{R}^{M \times T}$, where $M$ is the number of pixels and $T$ the number of frames in the time series. For simplicity,  we have $x$ and $t$ denote discretely sampled points in space and time in the notation, without further use of indices when the meaning is clear from the context. All data obtained with imaging instruments is discretely sampled. Any measurement of $d$ in the  $\tens{D}$ data can be described in the functional form evaluated at discrete points:
\begin{equation}
    \label{eq:signal_contributions}
    \begin{split}
    & D(x, t \,|\, N) =  S(x, t) + g(x, t \,|\, N) + \epsilon(x, t),
    \end{split}
\end{equation}
where $S$ is the planet signal contribution, $g$ denotes the functional form of the systematics affecting this measurement, and $\epsilon$ is the stochastic noise (e.g., photon noise). Here, $N$  is a stand-in for any hidden parameters that may be responsible for causing the systematics (e.g., wavefront phase residuals after adaptive optics (AO) correction, temperature, wind direction or speed).

In a data-driven approach to modeling the systematics, our goal is to determine $g(x, t \,|\, N)$ based on how  $g$ is expressed using other sets of data while being affected by the same underlying hidden parameters, $N$. If we could obtain a perfect model for $g$ simply by subtracting it, we could detect and determine the planet signal, $S$, up to a stochastic uncertainty term.
However, we have to guarantee that we do not incorporate contributions from $S$ into our systematics model, $g$, otherwise we will attenuate the signal of interest, an effect known as self-subtraction\footnote{If indeed the model is subtracted. Otherwise it constitutes a form of overfitting of the systematics model.}.
For this reason, we divide the data into two subsets of measurements. Those that are affected by signal from a planet above a chosen threshold\footnote{The threshold chosen does not matter for the formalism. In practice, however, it usually relates to some defined exclusion criterion based on the total flux overlap of a companion signal and training data. The companion PSF extends over the whole image at a very low flux level, therefore a meaningful cut-off has to be chosen for defining the extent of the signal.} are thus called $\mathcal{Y}$ and those that are not affected are called $\mathcal{X}$. The union of both sets represents all sampled position and time combinations, $\mathcal{D}$, and both sets are disjoined.
\begin{equation}
    \begin{split}
        \mathcal{Y} &= \{(x, t) \in \mathcal{D} \, | \, S(x, t) > \text{thr}\}, \\ 
        \mathcal{X} &= \{(x, t) \in \mathcal{D} \, | \, S(x, t) \leq \text{thr}\},\\
        &\mathcal{X} \cup \mathcal{Y} = \mathcal{D}, \quad \mathcal{X} \cap \mathcal{Y} = \emptyset. 
    \end{split}
\end{equation}
The resulting notion is that $g(x, t \, | \, N)$ for all $(x, t) \in \mathcal{Y}$ can be approximated using a combination of elements in $\mathcal{X}$ that are affected by the same underlying causes of systematics, $N$. The most direct approach to this type of problem is to assume a linear regression model. However, since the data itself is drawn from a multi-dimensional space (space and time), a choice has to be made as to the set of basis vectors within which the linear regression is performed. Traditionally, image vectors (or, more often, vectors of sub-image regions) have been used as the basis of the linear regression (e.g., all LOCI and PCA variants  mentioned earlier on). For one-dimensional representations, we may use a simplified vector notation where, for example, image vectors can be denoted as $\vec{d}_{t}(x)$, that is, as a data vector of pixel values for a set of positions $x$ at the time $t$. The image space is an intuitive basis when we think of the speckle noise as a relatively stable (quasi-static) pattern that is part of the (coronagraphic) stellar  point spread function (PSF). The PSF is intuitively thought of as a spatial construct.

Let us now formulate, in general terms, the spatial approach as it is implemented in LOCI-like algorithms, in which we are trying to describe an image data vector in terms of other image vectors taken at a different time.
We start by defining image regions $\setPy^k$ of pixels that are significantly affected by $S(x_i,t_k)$ at a given $t_k$ as subsets of the complete image space of all pixels, $\setPy^k \subset \mathcal{P:}$
\begin{equation}
\label{eq:image_region}
    \setPy^k = \{  x_i \, | \, S(x_i, t_k) > \text{thr} \}, \quad k \in \{1, 2, \dots, T\}.
\end{equation}

This corresponds to the area in an image covered by the companion PSF above a certain flux threshold at a given time. Given that the signal position changes over time, we can now define sets of times $\setTx^k$ in the observation sequence for which the planet signal does not affect this image region:
\begin{equation}
    \begin{split}
    \setTx^k = \{ t_{l} \, | \, \setPy^k \cap \setPy^{l} = \emptyset, \quad k, \, l \in \{1, 2, \dots, T\} \}
    \end{split}
    \label{eq:sets_for_spatial_model}
.\end{equation}
This is the set of all times at which the companion has moved far enough to not overlap with the area it occupied at a certain time. From these times we can build a linear model that does not incorporate planet signal.
We can then say that the systematics and planet flux in a particular image region $\setPy^k$ and for any particular time $t_k$ can be estimated using a linear least squares regression solving
\begin{equation}
    \argmin_{\substack{\omega_k, \,\alpha_{k, l} \\ \forall l}} \sum\limits_{x \, \in \, \setPy^k} \left|\left|
    \vphantom{\sum\limits_{t_l \,\in\,\setTx^k}}
    \right.\right.
    \underbrace{
    \vphantom{\sum\limits_{t_l \,\in\,\setTx^k}}
    D(x, t_k)}_{\text{data}} - \underbrace{\left(\omega_k \, \hat{S}(x, t_k) + \sum\limits_{t_l \,\in\,\setTx^k} \alpha_{k, l} \, D(x, t_l)\right)}_{\text{planet signal + systematics model}} \left.\left.
    \vphantom{\sum\limits_{t_l \,\in\,\setTx^k}}
    \right|\right|^2,
    \label{eq:minimization_spatial}
\end{equation}
where $\omega_k$ corresponds to the contrast of the planet signal estimated at time, $t_k$, and $\alpha_{k, l}$ are the weight coefficients of the linear model at time, $t_k$, and $\hat{S}$ is a model of the planet signal. In the spatial representation the planet model takes the form of an image of the companion PSF, $\vec{\hat{s}}_{t_k}(x)$, with $x \in \setPy^k$.
This approach assumes that the speckle pattern is either sufficiently stable (up to a scaling factor) or recurrent over the course of the observation (e.g., finite probability of jitter to revisit the same position, recurring Strehl ratio, or observing conditions in general). In other words, we assume we have enough measurements for $\vec{d}_{t}(x \, | \, N)$ with $t \in \setTx^k$ and $x \in \setPy^k$ to probe the underlying distributions of unobserved causal factors, $N$, to allow us to reconstruct a specific instance of the systematics function $g$. In the spatial representation $g$ corresponds to the speckle pattern, $\vec{g}_{t_k}(x)$, in the image region.
To a large extent this assumption is well founded, as evidenced by the success of the LOCI and spatial PCA-based family of algorithms. The details of how the $\alpha$-coefficients are optimized and how the planet contrast, $\omega$, is finally estimated vary strongly depending on the implementation \citep[e.g.,][]{Lafreniere2007, Pueyo2012, Marois2014, Wahhaj2015} and this is one of the main distinguishing factors between the large number of published algorithms. Here, two closely related problems are the problem of collinearity and the problem of overfitting of the planet signal. Considering that many of the image vectors used in the linear model share common features, one solution is to use regularization \citep[e.g., damped LOCI,][]{Pueyo2012}. The most popular approach is to use a truncated set of principal component images instead of using the image vectors  $\vec{d}_{t_l}(x \, | \, N)$ directly as done above for Eq.~\ref{eq:minimization_spatial}. This representation in a lower dimensional space simultaneously reduces collinearity in the training data and prevents overfitting of the model.

The question of whether the planet signal is fitted simultaneously to the systematics model (as done in this work), fitted after subtracting a systematics model (as has been the case for most early works), or subtracted before fitting the systematics model \citep[e.g.,][]{Galicher2011} is another key difference in various implementations.
Additionally, implementations vary in how the geometry and size of the image regions are defined and, furthermore, they may use two differently defined regions based on whether the region is used for optimizing the $\alpha$-coefficients (usually larger than the actual region of interest) or subtracting the model, which is another way of reducing overfitting.
Lastly, we note that reference star differential imaging \citep[RDI, eg.,][]{Smith1984, Lafreniere2009, Gerard2016, Ruane2017, Xuan2018, Bohn2020} is also based on the above-described spatial systematics model formalism, with the difference that the reference images are taken from observations of different host stars, which are consequently lacking the same planet signal contribution, $S$, by default. This paper is only concerned with individual pupil-stabilized observations.

\subsection{Problems with spatial systematics models}
The above-described assumptions impose a high requirement on the overall stability of the instrument and observing conditions. In modern instruments, a large part of the systematics can indeed be said to be quasi-static and correlated on timescales of minutes to hours \citep[e.g.,][]{Milli2016, Goebel2018}, making this approach applicable to these stellar-noise contributions.

However, the main drawback of the spatial modeling approach is the need to prevent the contamination of the systematics model with an astrophysically relevant signal by using a temporal exclusion criterion (usually defined as a protection angle). Its selection is a trade-off between excluding more frames to prevent self-subtraction and losing information related to the speckle evolution due to the exclusion of highly correlated frames close in time.
As was already noted early on \citep{Marois2006}, this trade-off gets gradually worse the smaller the angular separation of a companion with respect to the central star is, because the same FoV rotation corresponds to smaller and smaller physical displacement on the detector between subsequent frames. At small separations, a large fraction of the observation sequence will be excluded from the training set (see Appendix~\ref{sec:temporal_exclusion_impact}).

At small separations, these exclusion criteria cover time spans of the order of the linear decorrelation timescale for relatively stable instrumental speckles \citep{Milli2016, Goebel2018}. Even at large separations, the temporal exclusion will still be on the order of minutes. This means that using a spatial approach makes it intrinsically difficult to model the turbulence-induced, short-lived speckles with an exponential decay timescale of a few seconds \citep[$\tau$=3.5s,][]{Milli2016}. A sudden change in the overall conditions or the state of the instrument is likewise difficult to account for. Traditionally, obtaining simultaneous training data for modeling the systematics requires spectral differential imaging \citep[SDI,][]{Rosenthal1996, Racine1999} or polarimetric differential imaging \citep[PDI,][]{Kuhn2001}.

In this work, however, we demonstrate that it is possible to apply a causal, temporal systematics model to monochromatic pupil-tracking observations, using simultaneous but non-local reference data. This allows us to avoid the harsh temporal exclusion criterion used for preventing self-subtraction by introducing a less strict spatial exclusion criterion of pixels affected by signal, which is solely determined by the total FoV rotation, independent of the angular separation between the star and exoplanet.

\section{TRAP: A causal, temporal systematics model for high-contrast imaging}
\label{sec:causal_modeling}
Analogous to the spatial approach as described in Eqs.~\ref{eq:image_region},~\ref{eq:sets_for_spatial_model},~and~\ref{eq:minimization_spatial}, we can define the temporal approach in which all image vectors at a specific time become time series vectors at a specific location. This is done by first constructing a subset of times at which a specific pixel, $x_i$, is affected by planetary signal and construct a reference set of other pixels, $x_j$, which, in the same subset of times, does not contain planet signal. This would be a one-to-one translation of LOCI to a temporal approach, one that is local in time rather than space. Let us first construct the set of pixels affected by the planet signal at any point in time $t$ of the observation sequence:

\begin{equation}
    \setPy = \cup_k \setPy^k.
\end{equation}
For each affected pixel, $x_i \in \setPy$, we construct a training set of unaffected other pixels $\setPx^i$:

\begin{equation}
    \begin{split}
    \setTy^i &= \{ t \, | \, S(x_i, t) > \text{thr} \}, &\forall x_i \in \setPy, \\
    \setPx^i &= \{ x_j \, | \, \setTy^i \cap \setTy^{j} = \emptyset\},  &i, \, j \in \{1, 2, \dots, M\},
    \end{split}
    \label{eq:sets_for_temporal_loci_model}
\end{equation}
and subsequently, we can build our model similar to Eq.~\ref{eq:minimization_spatial} for any $x_i \in \setPy$
\begin{equation}
    \argmin_{\substack{\omega_i, \,\alpha_{i, j} \\ \forall j}} \sum\limits_{t \,\in\, \setTy^i} \left|\left|
    \vphantom{\sum\limits_{j \,\in\,\settildePx}}
    \right.\right.
    \underbrace{
    \vphantom{\left(\omega \hat{S}(x_i, t) + \sum\limits_{j \,\in\,\settildePx} \alpha_j D(x_j, t)\right)}
    D(x_i, t)}_{\text{data}} - \underbrace{\left(\omega_i \, \hat{S}(x_i, t) + \sum\limits_{x_j \,\in\,\setPx^i} \alpha_{i, j} \, D(x_j, t)\right)}_{\text{planet + systematics model}} \left.\left.
    \vphantom{\sum\limits_{j \,\in\,\settildePx}}
    \right|\right|^2,
\label{eq:minimization_temporal}
\end{equation}
where the data to be fitted takes the form of a (partial) time series (light curve), $\vec{d}_{x_i}(t \, | \, N)$, of one pixel, $x_i \in \setPy$, for all times, $t \in \setTy^i$, at which the pixel is affected by a planet signal above a certain threshold. In the temporal representation the planet model takes the form of a positive transit light curve $\vec{\hat{s}}_{x_i}(t)$ for $t \in \setTy^i$.
The systematics $g$ in the temporal representation correspond to temporal trends, $\vec{g}_{x_i}(t)$, with $t \in \setTy^i$ and are reconstructed in a basis set of reference time series, $\vec{d}_{x_j}(t \, | \, N)$, that are unaffected by planet signal during the same time.
In the literature, this process is usually referred to as a de-trending of light curves. The fitting of the systematics model can therefore be achieved in a mathematically analogous way to the spatial model when swapping the spatial and temporal axes.
However, in our implementation, we loosen the restriction on temporal locality, such that we always look at the complete time series and are left with only two relevant sets: the set of pixels that are at any point in time affected by planet signal $\setPy$ and those that are at no point affected $\setPx:$
\begin{equation}
    \begin{split}
    \setPy &= \{ x \, | \, \exists \, t \, \text{with}\, (x, t) \in \mathcal{Y} \} \\
    \setPx &= \{ x \, | \, \forall \, t \, \text{with}\, (x, t) \in \mathcal{X} \} \\
    & \setPx \cup \setPy = \mathcal{P}, \quad \setPx \cap \setPy = \emptyset,
    \end{split}
    \label{eq:sets_for_temporal_model}
\end{equation}
and the model is defined over all times, $t \in \mathcal{T,}$ such that for any \mbox{$x_i \in \setPy$}, Eq.~\ref{eq:minimization_temporal} becomes:
\begin{equation}
    \argmin_{\substack{\omega_i, \,\alpha_{i, j} \\ \forall j}} \sum\limits_{t \, \in \, \mathcal{T}} \left|\left|
    \vphantom{\sum\limits_{x_j \,\in\,\settildePx}}
    \right.\right.
    \underbrace{
    \vphantom{\left(\omega \hat{S}(x_i, t) + \sum\limits_{j \,\in\,\settildePx} \alpha_j D(x_j, t)\right)}
    D(x_i, t)}_{\text{data}} - \underbrace{\left(\omega_i \, \hat{S}(x_i, t) + \sum\limits_{x_j\, \in\, \setPx} \alpha_{i, j} \, D(x_j, t)\right)}_{\text{planet + systematics model}} \left.\left.
    \vphantom{\sum\limits_{x_j \,\in\,\settildePx}}
    \right|\right|^2.
\label{eq:minimization_trap}
\end{equation}
In an intuitive sense, it may not be obvious why we can use non-local time series to create a systematics model for the time series of a specific pixel, unless we consider the causal structure of the systematics that confounds the signal in the data.
While pixels physically separated enough on the detector do not directly ``talk to each other'' and are, at first order, independent measuring devices, they are influenced by the same underlying causes that generate the systematics in the first place. At the most basic level, the primary cause of systematics, that is, the contamination by the stellar signal, is the (coronagraphic) PSF of the star itself which changes with observing conditions and changes in the instrument. Of course, depending on the detailed underlying cause, the area of shared influence can be spatially different; for example, wind driven halo effects \citep{Cantalloube2018, Cantalloube2020} cannot be statistically inferred from reference pixels unaffected by the same underlying cause.

A detailed mathematical description of modeling systematics in time series data based on shared underlying causes can be found in \citet{Schoelkopf2016}, giving the statistical background and justification for Eqs.~\ref{eq:minimization_temporal}~and~\ref{eq:minimization_trap}.
The above-mentioned work shows that by using a regression model of a sufficiently large number of ``half-siblings,''  that is, measurements that do not share the signal but the same causal relationship with the systematics, $\vec{d}_{x_j}(t \, | \, N)$, of one pixel, $x_j \in \setPx$, we can reconstruct a specific instance of the systematics function, $\vec{g}_{x_i}(t \, | \, N)$. Furthermore, we can include a model of the planet signal itself in the regression, such that the overall regression explains the data in $\setPy{}$ up to a stochastic noise term (see Eq.~\ref{eq:signal_contributions}), which remains the fundamental limit of the data. This fundamental limit is a combination of photon noise, detector read-out noise, and flat fielding uncertainties. In practice, we will still be limited by imperfections in the systematics model because only a finite number of regressor light curves are available and some causes of systematics may not be perfectly shared among them.\\

The mathematical idea of using a causal time-series regression model, as described above, has been employed very successfully for transit observation using the \textit{Kepler} spacecraft \citep[e.g.,][]{Wang2016a}. The benefit of simultaneously including a forward model of the expected transit shape in the regression model has been demonstrated in detail in \citet{Mackey2015}.

\subsection{Implementation and application to high-contrast imaging data}
\label{sec:algorithm}
The situation is very similar for high-contrast imaging data. The causes of systematics, with the exception of detector artifacts (e.g., bad, warm, or dead pixels, flat field), usually influence either the image globally (e.g., most atmospheric effects, Strehl ratio variations), or a significant region of the detector (e.g., wind-driven halo, (mis-)alignment of the coronagraph). Even slowly evolving changes in the quasi-static speckle pattern caused by the instrument usually are not strictly confined to one region of the detector (e.g., speckle patterns can and do indeed display point symmetries).

\begin{figure}[t]
  \centering
  \includegraphics[width=\columnwidth]{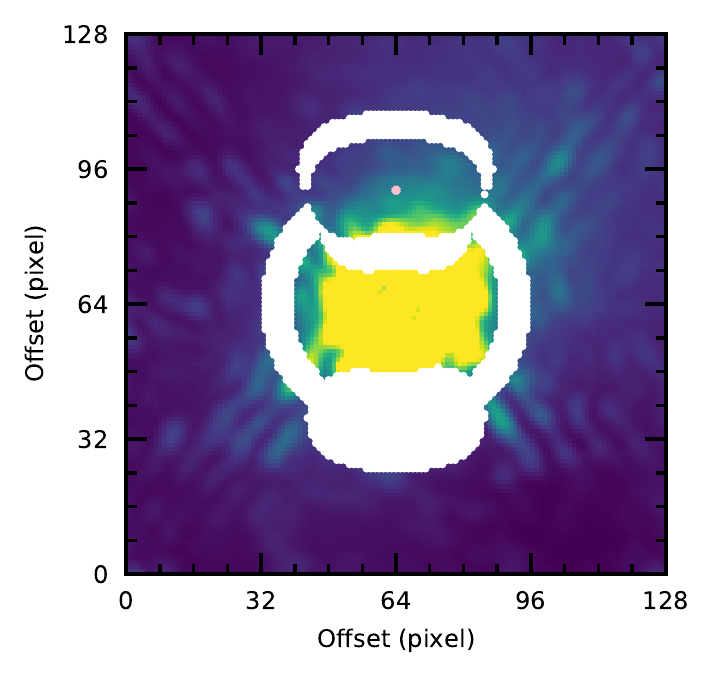}
  \caption[Reference pixel selection in TRAP]{Example of reference pixel selection for one assumed planet position. The signal here is assumed to move through the position $(\Delta x, \Delta y) \approx (0, 28)$ pixel above the star at the midpoint of the observation sequence. This position is marked by a pink cross and represents the center of the reduction area, $\setPy$. The reference pixels, $\setPx$, are shown in white. This example is based on the 51~Eri observation's parallactic angles, the annulus is seven pixels wide.}
  \label{fig:regressor_selection}
\end{figure}

A detailed study of the objectively optimal choice of reference pixels to capture most shared underlying systematic causes is beyond the scope of this work, but multiple heuristic choices can be made, namely: 1) a preference for pixels at a similar separation \citep[same underlying modified Rician speckle intensity distribution,][]{Marois2008}; 2) pixels at same position angle inwards and outwards of the reduction area \citep[position angle-dependent effects, e.g., wind-driven halo,][]{Cantalloube2018, Cantalloube2020}; 3) inclusion of all pixels analogous to the reduction region but on the opposite side of the star \citep[point symmetric features arising from the nature of phase-aberrations, see e.g. small-phase approximation,][]{Perrin2003, Ren2012, Dou2015}; and 4) exclusion of all pixels affected by the planet signal. Varying the regressor-selection parameters around these heuristics does not significantly impact our final results. Figure~\ref{fig:regressor_selection} shows an example for the reference pixel (regressor) selection geometry chosen in this work for an assumed companion that passes north of the host star at the midpoint of the observation. The annulus used has a width of 7 pixels and the PSF used includes the first Airy ring and has a diameter of 21 pixels.\\

In our implementation\footnote{\url{https://github.com/m-samland/trap}}, called the Temporal Reference Analysis of Planets (TRAP), we use: a)~a linear model with a quadratic objective function (assuming Gaussian uncertainties\footnote{This is a simplified assumption and reality is more complex in terms of the underlying distributions \citep[e.g.,][]{Marois2008, Pairet2019}. Additionally, noise is generally heteroscedastic and correlated. We mitigate this issue by empirically rescaling the derived uncertainties at the end.}); b)~the principal components of the regressor pixel light curves, not the pixels themselves (principal component regression); c)~a fit of each affected pixel individually instead of all affected pixels at the same time for computational reasons; d)~a fit of the planet model and the systematics model simultaneously in order to prevent overfitting, similar to the approach by \citet{Mackey2015}. The above choice of using principal components reduces the collinearity of the regressors used for the systematics model and transforms our model onto an orthogonal basis. It also allows us to forgo regularization in favor of truncating the principle components after a certain number. 
Furthermore, we do not use any spatial pre-filtering of the data as is commonly done in many implementations to remove structures correlated on larger scales than the expected planet signal from the data \citep[e.g., as done in][]{Cantalloube2015, Ruffio2017}. We also do not pre-select or discard any frames as bad frames in this study.\\

Given a data cube of the observation sequence, $\tens{D} = d_{ik} \in \mathbb{R}^{M \times T}$, where $M$ is the number of pixels and $T$ is the number of frames in the time series, as well as the known parallactic angles for each frame, we know the temporal evolution of a companion signal's position. The algorithm is applied as follows for one assumed relative position $\theta$ $(\Delta \text{RA}, \Delta \text{DEC}$) of a companion point source on-sky and its corresponding trajectory on the detector:
\begin{enumerate}
    \item We generate a forward model of the planet signal at an assumed position, $\theta$, for the time series by embedding the model PSF in an empty data cube at the appropriate parallactic angle for each exposure to obtain the set of pixels $\setPy$ affected at any point in time during the observation. Conversely, all pixels, $\setPx$, not affected at any time by planet signal are potential reference pixels.
        
    Additionally we obtain the corresponding planet model light curve $\vec{\hat{s}}_{x_i}(t)$ for every $x_i \in \setPy$. We use the unsaturated PSFs obtained without the coronagraph directly before or after the sequence, but artificially induced satellite spots could also be used. The PSF has been adjusted to the exposure time and filter of the science exposures and is set to zero beyond the first Airy ring.
        
    \item Instead of using all time series from unaffected pixels, $\setPx$, to build the systematics model, we construct the training set with additional desired constraints $\settildePx \subset \setPx$ as previously described (see Fig.~\ref{fig:regressor_selection}). The number of reference pixels in the reference set is called $R = | \settildePx |$.

    \item For each pixel $x_i \in \setPy$:
    \begin{enumerate}
        \item We want to simultaneously optimize the coefficients of our systematics and planet model.
        
        We first show how this done starting from Eq.~\ref{eq:minimization_trap}, which uses the reference light curves from the data directly. The minimization problem can be solved as a system of linear equations of the form:
        \begin{equation}
            \label{eq:linear_regression}
                \tens{A_{\theta, i}}^T \tens{C_i}^{-1} \tens{A_{\theta, i}} \, \vec{c}_i = \tens{A_{\theta, i}}^T \tens{C_i}^{-1} \vec{b}_i
        \end{equation}
        because Eq.~\ref{eq:minimization_trap} is linear in $\omega_i$ and $\alpha_{i, j}$. $\tens{A_{\theta, i}}$ is the design matrix, also called regressor matrix, containing the light-curve vectors constituting the terms of the linear model to be fitted. Here, $\vec{c}_i = (\alpha_{i,0}, \dots, \alpha_{i,R}, \omega_i)^T$ is the vector of model coefficients, $\vec{b}_i = \vec{d}_{x_i}(t)$ is the light-curve data vector of pixel, $x_i$, that is to be fitted, and $\tens{C_i}$ is the covariance  associated with the data vector to be fitted. The uncertainties of the time series are in general heteroscedastic (not uniform in time), but uncorrelated, such that the off-diagonal elements are zero.
        
        \item The design matrix, $\tens{A_{\theta, i}}$, consists of columns containing the additive terms of the overall model we want to fit. As such the part of the design matrix for Eq.~\ref{eq:minimization_trap} that describes only the systematics model consists of columns that contain the reference light curves, $d_{kj}={d_{jk}}^{T}$, such that $j \, | \, x_j \in \settildePx$, $j \in \left\{1,2,\dots, R\right\}$, and $k \in \left\{1, 2, \dots, T\right\}$. The same basis vectors describing the systematics model are used for all pixels $x_i$ for a given planet position $\theta$.

        Since this would only account for the systematics, we add two additional columns, one containing a constant offset to be fitted and one containing the light curve vector, $\hat{\vec{s}}_{x_i}(t)$, describing the companion signal at pixel, $x_i$, over time\footnote{Here, it is possible to include additional sources of information on temporal behavior as columns, such as auxiliary data on atmospheric conditions.}. This (preliminary) design matrix~$\tens{A}_{\theta, i}$ looks as follows:
        \begin{equation}
        \label{eq:design_matrix_1}
        \tens{A}_{\theta, i;\, kj} =
            \begin{cases}
              d_{kj}, & \text{if}\ j \leq R\\
              1, & \text{if}\ j = R+1 \\
              \hat{S}_{k,i}, & \text{if}\ j = R+2
            \end{cases}
            \: \in \mathbb{R}^{T\times (R+2)}.
        \end{equation}
        
        \item However, this results in an overdetermined and highly collinear problem because there are typically more reference light curves than there are samples in time ($R \gg T$) and the light curve share many common features. Therefore, some form of regularization or dimensionality reduction is required.
        We opt for representing them in a lower dimensional space using the principal components of the light curves instead of the reference light curves themselves. This is implemented using a singular value decomposition (SVD)\footnote{In practice, we use the compact SVD to improve computational performance. We note that the SVD can become ill-defined for the inner-most regions, when the temporal sampling of the data is very high. In this case, we should ensure that the number of regressor pixels is not too small compared to the number of temporal samples.}. We first center the data robustly by subtracting the temporal median of each light curve $d_{kj}$. 
        This centered data matrix can then be factorized into $\tens{U} \, \Sigma_\textrm{SVD} \, \tens{V}^{T}$, where $\tens{U}$ corresponds to a $T \, \times \,T$ matrix, $\tens{\Sigma_\textrm{SVD}}$ corresponds to a $T \, \times \,R$ diagonal matrix with non-negative real numbers, and $V^{T}$ corresponds to an $R \, \times R$ matrix. Because the data $d_{kj}$ is real, $\tens{U}$ and $\tens{V}^{T}$ are real and orthonormal matrices. In the following context, we use $\tens{U}$, the columns of which are called left-singular vectors and form an orthonormal basis in which the light curves, $d_{kj}$, can be described. They are ordered in decreasing order of explained variance. We truncate the number of components to be fitted to remove spurious signals from the systematics model and reduce overfitting. The number of components used in the algorithm is specified by the fraction, $f$, of the maximum number of components, $N_\mathrm{max} = T$, such that $N_\mathrm{used}=f\,N_\mathrm{max}$ with $f \in [0, 1]$ and $N_\mathrm{used} \in \mathbb{N}$. $N_\mathrm{used}$ is always rounded to the closest integer for any chosen $f$. The fraction of components, $f$, is a user parameter that can be set. The regressors for the temporal systematics are then given by $\tens{U}_{kl}$, where $l \in \left\{1,2,\dots, N_\mathrm{used}\right\}$. The final design matrix is:
        \begin{equation}
        \label{eq:design_matrix_2}
        \widetilde{\tens{A}}_{\theta,i;\, kl} =
            \begin{cases}
              \tens{U}_{kl}, & \text{if}\ l \leq f \, N_\mathrm{max}\\
              1, & \text{if}\ l = f \, N_\mathrm{max}+1 \\
              \hat{S}_{k,i}, & \text{if}\ l = f \, N_\mathrm{max}+2
            \end{cases}
            \: \in \mathbb{R}^{T\times (f \, N_\mathrm{max}+2)}.
        \end{equation}

        \item The vectors of the design matrix are mostly orthogonal and we can robustly invert the equation analogously to Eq.~\ref{eq:linear_regression} using linear algebra operations to obtain the vector of model coefficients and their associated uncertainty. The coefficients and their associated covariances are given by:
        \begin{equation}
        \begin{split}
            \vec{c}_i &= [\widetilde{\tens{A}}_{\theta,i}^T \, \tens{C_i}^{-1} \, \widetilde{\tens{A}}_{\theta, i}]^{-1}\, [\widetilde{\tens{A}}_{\theta,i}^T \,\tens{C_i}^{-1} \, \vec{d}_{x_i}] \\
            \tens{\Sigma}_{i} &= [\widetilde{\tens{A}}_{\theta,i}^T \, \tens{C_i}^{-1} \, \widetilde{\tens{A}}_{\theta,i}]^{-1}.
        \end{split}
        \label{eq:least_squares}
        \end{equation}
        In our current implementation, the covariance matrix of the data $\tens{C}$ is assumed to be uncorrelated in time (zero for off-diagonal elements), as the residuals after fitting the model are approximately white. The variance for the coefficients are therefore given by:
        \begin{equation}
            \sigma_{i}^{2} = \textrm{diag}(\tens{\Sigma_i}).
        \end{equation}
        The results shown throughout the paper assume identically distributed data uncertainties. However, known uncertainties for each pixel (e.g., photon noise, read-noise) can be passed to the routine, which is generally recommended if the variance of the input data is known. Taking into account photon noise puts less weight on data with higher speckle noise (e.g., pixels closer to the star, located on the spider, frames taken under worse conditions), further increasing the robustness of the reduction.
        The elements $\omega_i=c_{i,\,N_\mathrm{used}+2}$ and $\sigma_{\omega,i}^2=\sigma_{i,\,N_\mathrm{used}+2}^2$ correspond to the contrast of the companion and its respective variance by construction. A graphical example for the system of equations in Eq.~\ref{eq:linear_regression} to be solved is shown in Fig.~\ref{fig:system_of_equations}. The result of the fit of the time series are shown in Fig.~\ref{fig:fitted_data}. The above example is for the central pixel of the signal-affected detector area shown in Fig.~\ref{fig:regressor_selection}, injected with a relatively bright signal ($10^{-4}$ contrast) for demonstration purposes.
    \end{enumerate}
    \item After obtaining $\omega_i$ and $\sigma_{\omega,i}^2$ for all $x_i \in \setPy$, we remove significant outliers whose median of the residual vector deviates more than a set threshold in robust standard deviations from the rest (in our implementation 3 robust standard deviations). This usually removes a low single-digit number of pixels (<10), consistent with the number of anomalous (e.g., bad or cosmic ray affected) pixels that are expected to be present in an area the size of a typical reduction region. The incurred information loss is minimal. We call this new subset of pixels: $\settildePy \subset \setPy$.
    \item Finally, we take the average of the planet contrast coefficients weighted by their respective uncertainties over all remaining pixels:
        \begin{equation}
            \label{eq:weighted_average}
        \omega_\theta = \frac{\sum\limits_{i}{\omega_i \, \sigma_{\omega,i}^{-2}}}{\sum\limits_{i}{\sigma_{\omega,i}^{-2}}}, \quad
        \widetilde{\sigma}_\theta = \sqrt{\left(\sum\limits_{i}{\sigma_{\omega,i}^{-2}} \right)^{-1}}, \quad i \, | \, x_i \in \settildePy,
    \end{equation}
    to obtain a single contrast and preliminary uncertainty value for this on-sky position $\theta$. Equation~\ref{eq:weighted_average} assumes the pixel residuals to be uncorrelated in space, which, in practice, is not the case. The preliminary uncertainty, $\widetilde{\sigma}_\theta$, needs to be scaled with an empirical normalization factor, the derivation of which is discussed below, to account for simplified assumptions about the covariance structure of the data and spatial correlations between the residuals.
\end{enumerate}
Iterating over a grid of possible positions\footnote{We use the same spatial sampling as the pixel scale of the instrument.} allows us to construct a contrast and pseudo-uncertainty map, that is, the contrast $\omega_\theta$ and its preliminary uncertainty $\widetilde{\sigma}_\theta$ given the positions of the assumed companion relative to the central star. From these we construct a pseudo-signal-to-noise (S/N) map. The above uncertainties are computed under the simplified assumption that all measurements are Gaussian and independent, which may not accurately reflect the reality of high-contrast imaging data. Multiple studies have shown that residuals after post-processing with high-contrast imaging pipelines, while they may be significantly whitened, are not strictly Gaussian and independent \citep{Marois2008, Cantalloube2015, Pairet2019}. The most direct solution would be to account for these effects in the likelihood function used, but in practice, this has proven to be challenging since the uncertainty distributions are heteroscedastic and depend on the observing conditions, instrument, location in the image, and time. Therefore, we apply the same solution to this problem as in ANDROMEDA \citep{Cantalloube2015}, and, subsequently, in FMMF \citep[forward model matched filter, see ][]{Ruffio2017} and we empirically normalize the pseudo-S/N map using the azimuthal robust standard deviation of the S/N map itself computed in three-pixel-wide annuli as a function of separation. We denote these final uncertainty values as:
\begin{equation}
    \sigma_\theta = \widetilde{\sigma}_\theta \,  \sigma_{\mathrm{annulus},\theta}.
\end{equation}

Any reference to S/N in this work refers to the normalized S/N $\omega_\theta / \sigma_\theta$. This normalized S/N\ map is also referred to as the detection map. Any a priori known companion objects or background star sources will be masked out when deriving the normalization to avoid biasing the result. 

\begin{figure}
    \centering
    \includegraphics[width=\columnwidth]{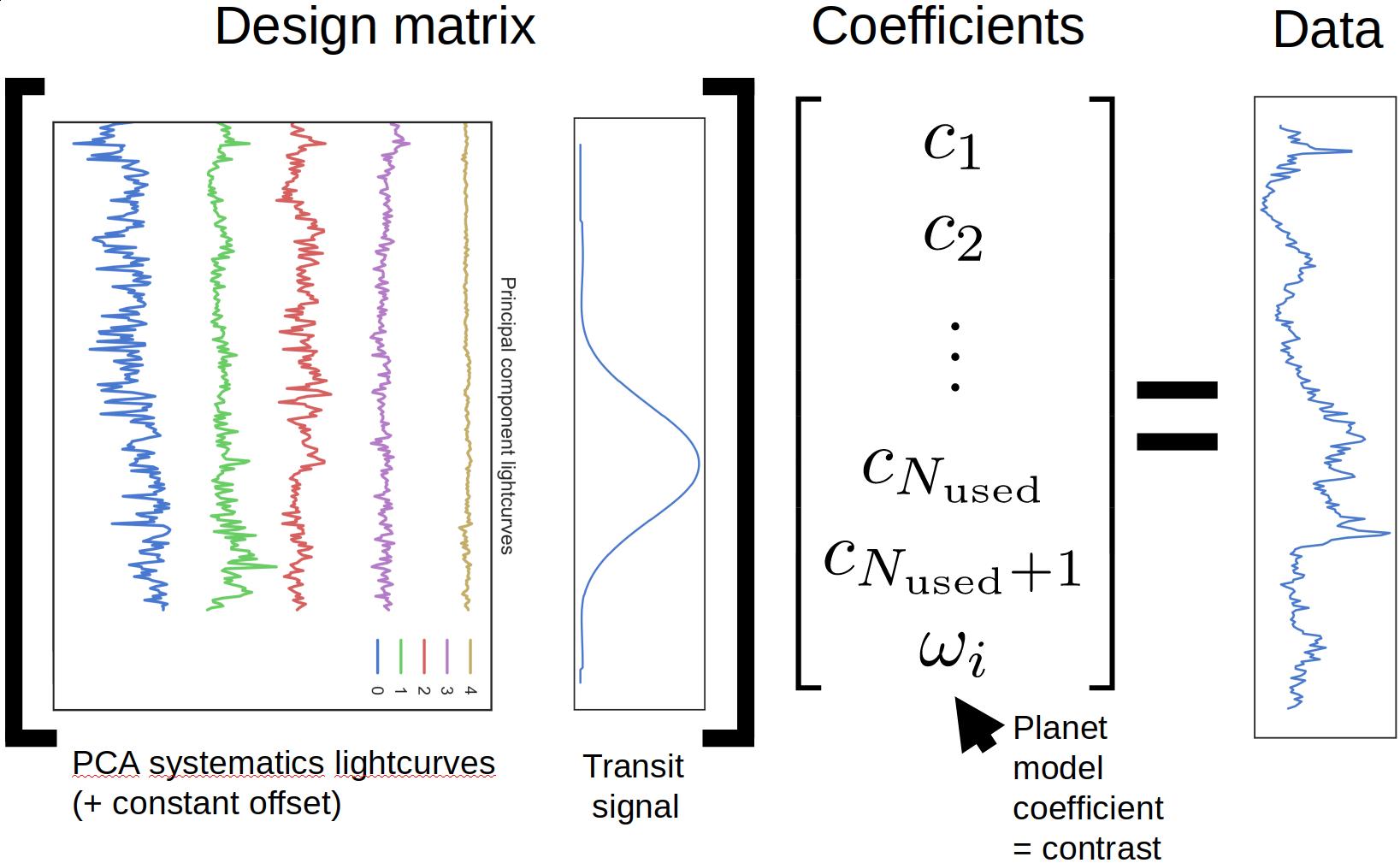}
    \caption{Example for Eq.~\ref{eq:linear_regression}, showing the first five principal component light curves and the planet model. The constant offset term ($c_{N_\text{used}+1}$) is not shown in the design matrix. This example shows a pixel with a bright ($10^{-4}$ contrast) injected planet signal based on 51~Eri~b data. The data corresponds to the central pixel in the injected planet's trajectory shown in Fig.~\ref{fig:regressor_selection} at the location marked by the pink cross. The principal components were determined from the reference pixels, $\setPx$, shown in white.}
    \label{fig:system_of_equations}
\end{figure}

\begin{figure}
        \centering
        \includegraphics[width=\columnwidth]{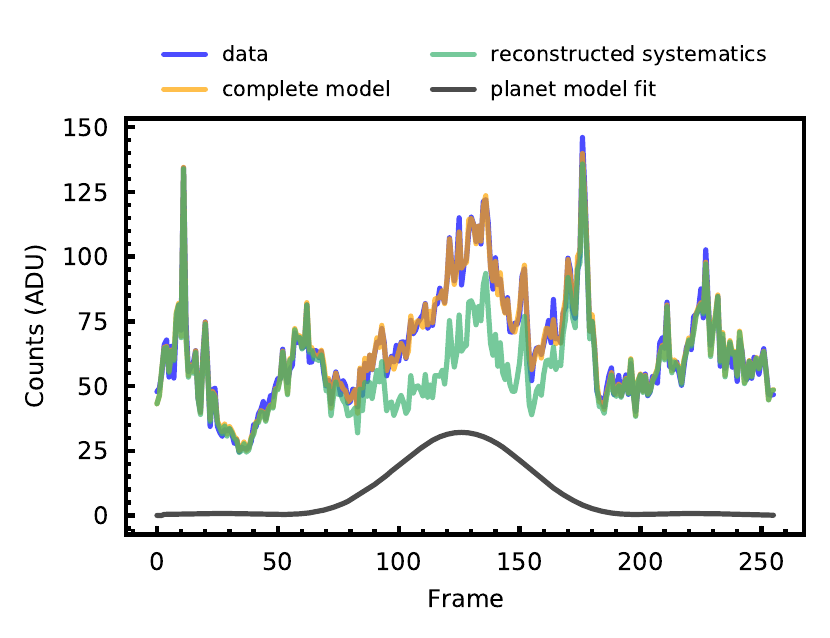}
        \caption[Time series fitted for planet signal and systematics]{Result of the fit for the pixel shown in Fig.~\ref{fig:system_of_equations}, with the reconstructed model for the time series with and without including the coefficient $\omega_i$ describing the amplitude of the companion signal.}
        \label{fig:fitted_data}
\end{figure}

\section{Datasets used for demonstration}
\label{sec:data}
\begin{figure*}[t]
        \centering
        \includegraphics[width=\textwidth]{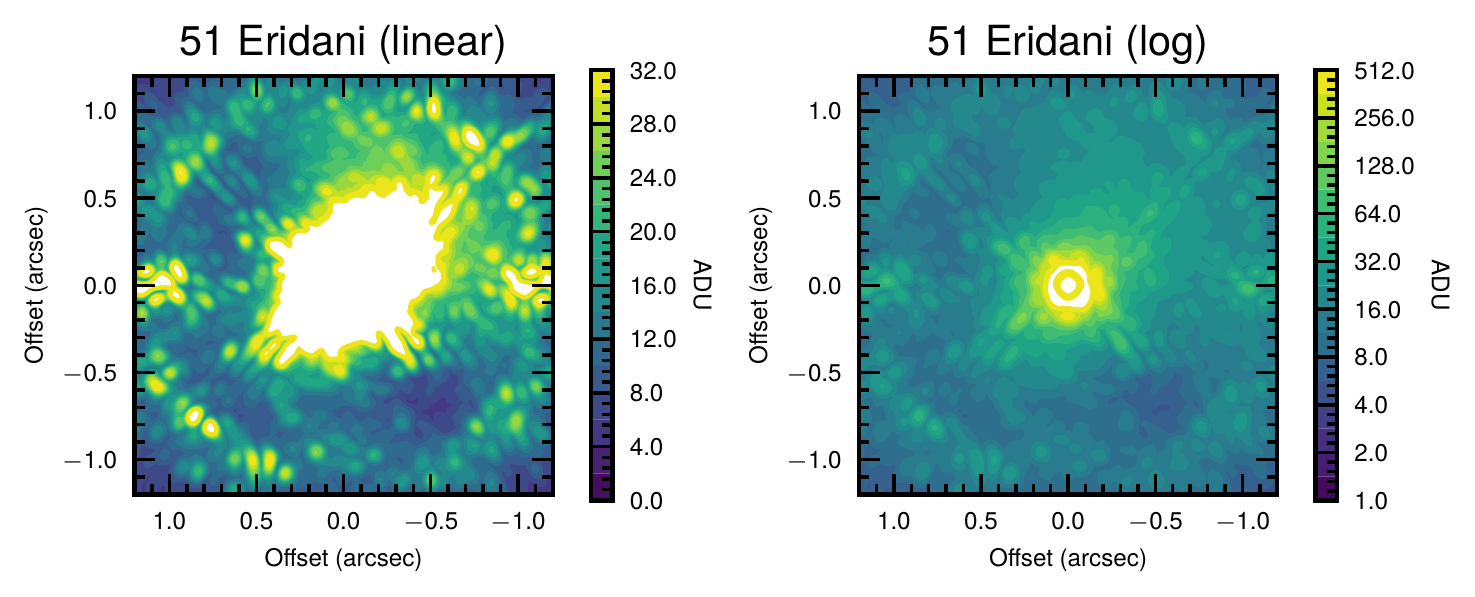}
        \caption[51~Eri pre-processed median combined image]{Contour map of median combined image of pre-processed 51 Eridani data cube. The left panel shows data in linear scaled brightness bins, the right panel in logarithmic scaling.}
        \label{fig:51eri_image}
\end{figure*}

\begin{figure*}[t]
        \centering
        \includegraphics[width=\textwidth]{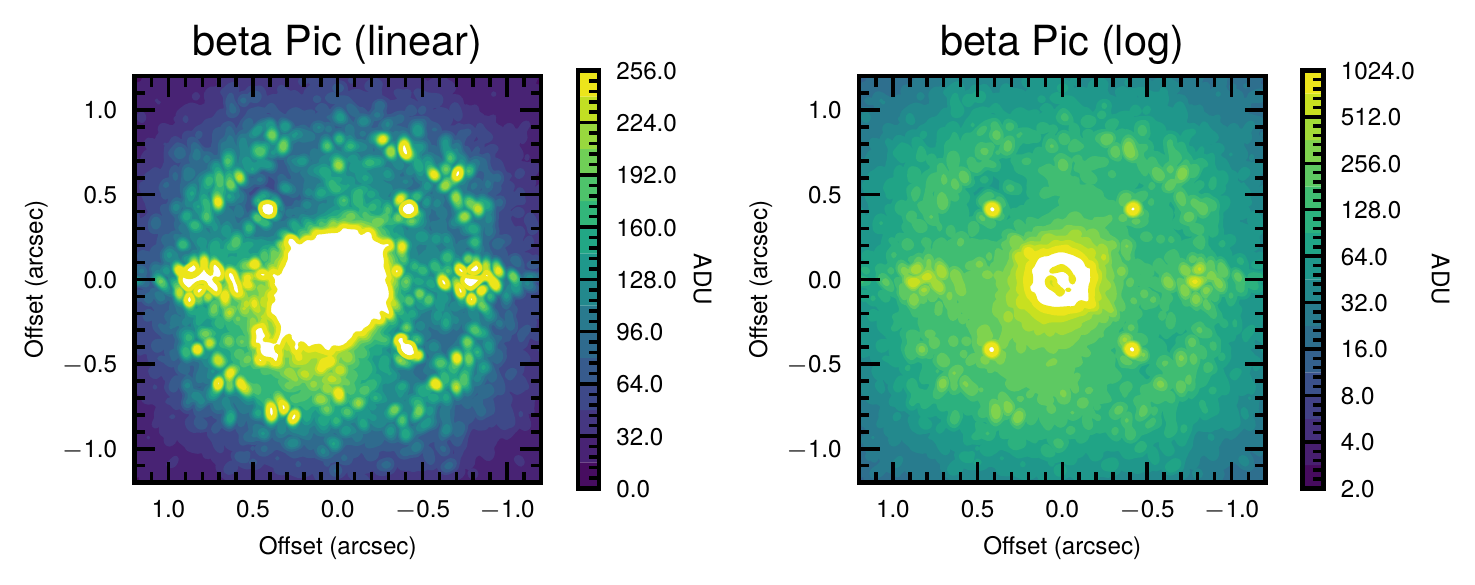}
        \caption[$\beta$ Pic pre-processed median combined image]{Contour map of median combined image of pre-processed beta Pic data cube. The left panel shows data in linear scaled brightness bins, the right panel in logarithmic scaling.}
        \label{fig:betapic_image}
\end{figure*}

We tested our TRAP algorithm on two real datasets obtained with the extreme-AO \citep[SAXO;][]{Fusco2014} fed infrared dual-band imager \citep[IRDIS;][]{Dohlen2008} of SPHERE \citep{Beuzit2019} at the VLT. Both datasets were obtained as part of the SHINE \citep[SpHere INfrared survey for Exoplanets,][]{Chauvin2017b, Vigan2020arXiv} large survey. The observation were obtained in pupil-tracking mode and use the apodized Lyot coronagraph \citep{Soummer2005} with a focal-mask diameter of 185 milli-arcsec. The two datasets are described in Table~\ref{tab:paper2_observations} and Figs.~\ref{fig:51eri_image}~and~\ref{fig:betapic_image} show the pre-processed temporal median image for the \object{51 Eridani} and \object{$\beta$~Pictoris} observations, respectively.

Both datasets were pre-reduced using the SPHERE Data Center pipeline \citep{Delorme2017DC}, which uses Data Reduction and Handling software \citep[DRH, v0.15.0,][]{Pavlov2008} and additional custom routines. It corrects for detector effects (dark, flat, bad pixels), instrument distortion, in addition to aligning the frames and calibrating the relative flux.

The first test case is the directly imaged planet in the 51~Eridani system \citep{Macintosh2015}. The data we use was previously published in \citet{Samland2017}. This dataset was obtained in the K1 and K2 bands simultaneously, but we focus our discussion on the K1 channel in which the companion is visible. The sequence was not taken under ideal conditions. It exhibits common problems encountered in high-contrast imaging, such as a strong wind-driven halo, as seen in Fig.~\ref{fig:51eri_image}.
These effects make for a realistic test case, and allow us to demonstrate the algorithm performs well under adverse and changing conditions.

The second test case uses IRDIS H2 data of the $\beta$~Pic system \citep{Lagrange2009} published in \citet{Lagrange2019} taken in continuous satellite spot mode, that is, using sine-wave modulations on the deformable mirror to induce four satellite spots in all frames of the sequence that can be used to determine the center accurately. This also permits a measurement of the satellite spot amplitude variations as proxy for the planet PSF model's flux modulation over time.
The purpose behind including this dataset is two-fold. Firstly, the exposure time is relatively short for an IRDIS sequence ($4\,$s), which allows us to test our time-domain based algorithm on a dataset with better temporal sampling, approaching the half-life time of fast-decaying speckles ($\tau=3.5\,$s), such that multiple correlation timescales are not averaged. Secondly, datasets taken in continuous satellite spot mode do not use detector-stage dithering, which is generally used for IRDIS coronagraphic sequences. We will use this non-dithered dataset to test our algorithm directly on the non-aligned data cube (only dark and flat corrected). This is only possible in a modeling framework that does not optimize the local spatial similarity between training and test sets, which is the case for our temporal approach. We demonstrate that we can skip bad pixel interpolation and shifting steps by including the shift in the forward model of the planet signal, and exclude the stationary bad pixels from our training and test sets altogether. This makes the propagation of all uncertainties throughout the complete data reduction feasible because the data does not need to be resampled or interpolated. We also account for SPHERE's anamorphism in the relative position of the companion model. 
We further demonstrate the possibility of including the satellite spot modulations in the forward model to reduce a systematic bias of the photometric calibration.

\begin{table*}
\small
\caption[Data used for TRAP]{Observing log}
\centering
\begin{tabular}{l c c c c c c c c c}
\hline\hline
Target & UT date& Satellite spots\tablefootmark{a} & IRDIS Filter & IRDIS DIT\tablefootmark{b}& $\text{T}_\text{exp}$&Field Rot.\tablefootmark{c}& Sr\tablefootmark{d} & DIMM seeing & $\tau_0$\\
& & & & (sec, \#) &(min)&(deg)& &($^\circ$)&(ms)\\
\hline
51 Eri & 2015-09-25 & no & K12 & $16\times256$ & 68 & 42 & 0.80--0.90 & 0.75 & 5.7\\
$\beta$ Pictoris & 2015-11-30 & yes & H23 & $4\times 200$& 53 & 40 & N/A & 1.06 & 10.9\\
\hline
\end{tabular}
\label{tab:paper2_observations}
\tablefoot{\footnotesize
Observational data used for the analyses. Seeing and coherence time $(\tau_0)$ correspond to the mean of the coronagraphic observation sequence. \tablefoottext{a}{Continuous satellite spot mode: calibration spots are always present in the coronagraphic data.}
\tablefoottext{b}{Detector integration time.}
\tablefoottext{c}{All observation were centered on the meridian passage of the target with an airmass between 1.08 and 1.13.}
\tablefoottext{d}{Strehl ratio computed by AO system's Real Time Computer, and scaled to a wavelength of 1.6 $\muup$m.}
}
\end{table*}

\section{Results}
\label{sec:paper2_results}
We performed a direct comparison of results between TRAP and the official IDL implementation of the ANDROMEDA algorithm \citep{Cantalloube2015}. It is always challenging to directly compare a new approach with existing pipelines due to numerous differences in implementation, a wide range of possible reduction parameters, test data taken in various conditions and observing modes, as well as differences in statistical evaluation of the outputs. We note that no comparison will ever be complete. For our work, we have chosen ANDROMEDA as the most viable representation of the spatial systematics modeling class of algorithms, because both TRAP and ANDROMEDA implement a likelihood-based forward modeling approach of the companion signal. This allows us to directly evaluate and compare the outputs in a fair way, such that differences in implementation do not impact the statistical interpretation of the results. Additionally, ANDROMEDA has been shown to compare well with other established PCA and LOCI-based pipelines \citep[e.g.,][]{Cantalloube2015, Samland2017}.

For our tests and comparisons, we use the exact same normalization method for the detection map and contrast curve computation for both TRAP and ANDROMEDA outputs. For the computation of the empirical normalization and the contrast curves, we mask the position of the planet with a mask size radius of $r_\text{mask} = 15$ pixels in the output maps ($\sim$180 mas).

We show the results for TRAP for a range of principal component fractions, $f$, (see Eq.~\ref{eq:design_matrix_2}). Our representative choice is $f=0.3$, which does not significantly underfit nor overfit either dataset in our reductions. The impact of $f$ is explored in more detailed in Section~\ref{sec:principal_components}. For ANDROMEDA, the representative choice of protection angle (minimum angular displacement of companion signal between two frames) is $\delta=0.5\,\lambda / D$, which shows good performance for extreme-AO SPHERE data and was the final parameter choice used for the spectro-photometric analysis of 51~Eri~b \citep{Samland2017}. The contrast curve for $\delta=1.0 \,\lambda / D$ is included for demonstrating the impact of the protection angle on contrast performance at small angular separations. Contrast curves for ANDROMEDA using $\delta=0.3,\,0.5,\,1.0\,\lambda / D$ for both datasets are shown in Appendix~\ref{sec:andromeda_protection_angle} for completeness.

\subsection{51 Eridani b: centrally aligned data cube}

\begin{figure*}[t]
        \centering
        \includegraphics[width=\textwidth]{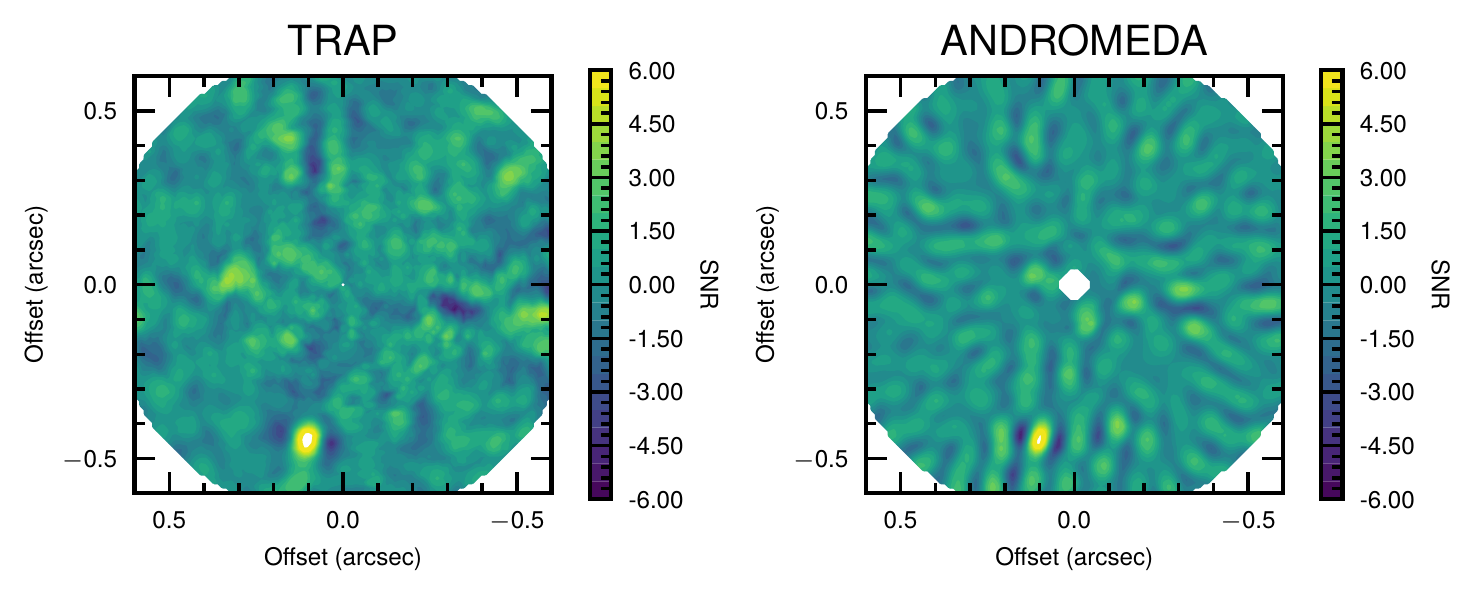}
        \caption[Detection map comparison between TRAP and ANDROMEDA for 51~Eri~b]{Contour map of normalized detection maps obtained with TRAP ($f=0.3$) and ANDROMEDA ($\delta = 0.5 \lambda/D$) for 51~Eri~b. These maps must not be confused with a derotated and stacked image. They represent the forward model result for a given relative planet position on the sky ($\Delta$RA, $\Delta$DEC), that is, the conditional flux of a point-source predicted by the forward model given a relative position, corresponding to a trajectory over the detector (all pixels affected during the observation sequence). The dark wings around the detection are not a result of self-subtraction, but purely a result of the fixed forward model position with respect to the real signal.}
        \label{fig:norm_detection_51eri}
\end{figure*}

\begin{figure*}
        \centering
        \begin{tabular}{cc}
        \includegraphics[width=.49\textwidth]{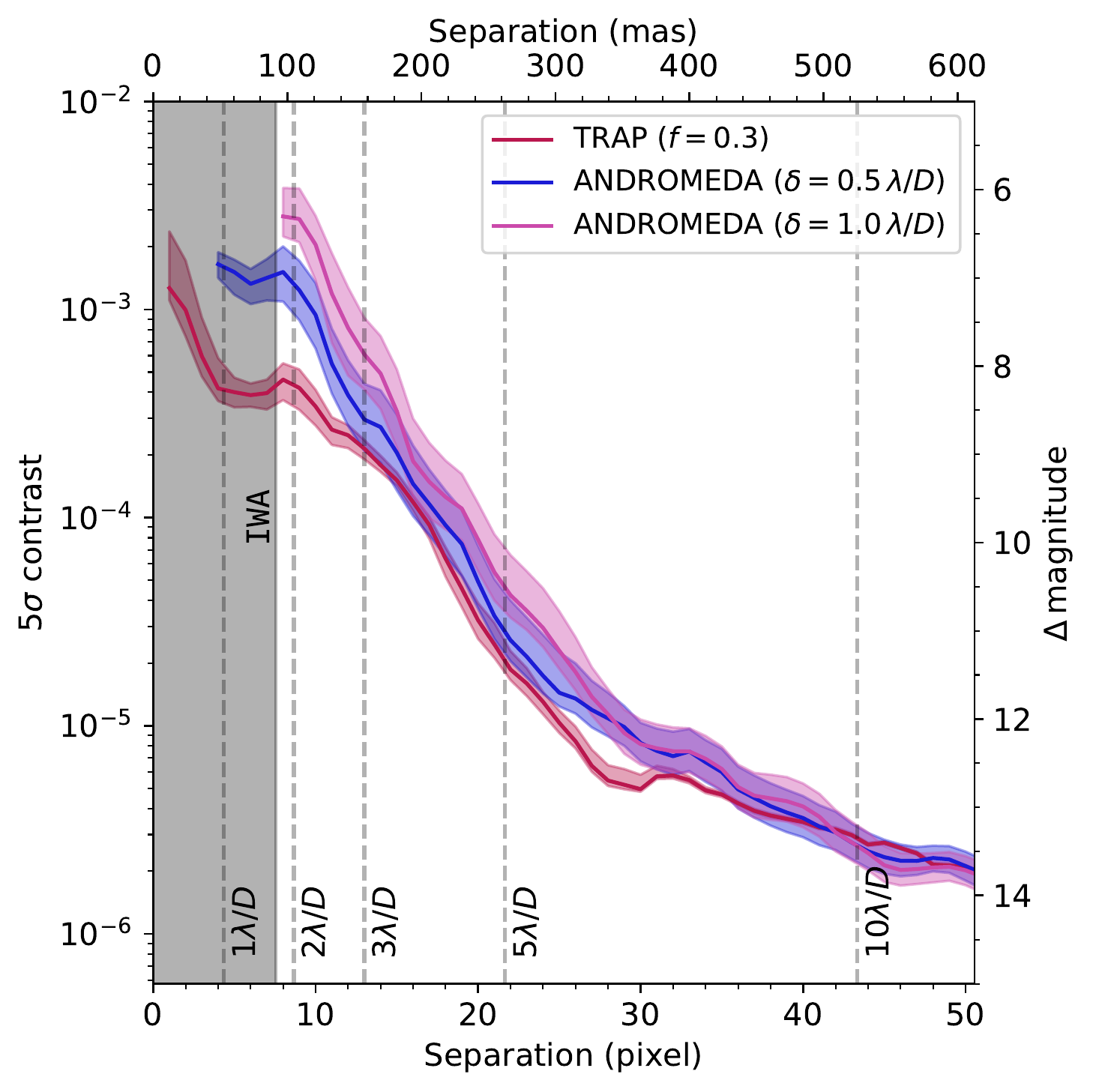}
        \includegraphics[width=.49\textwidth]{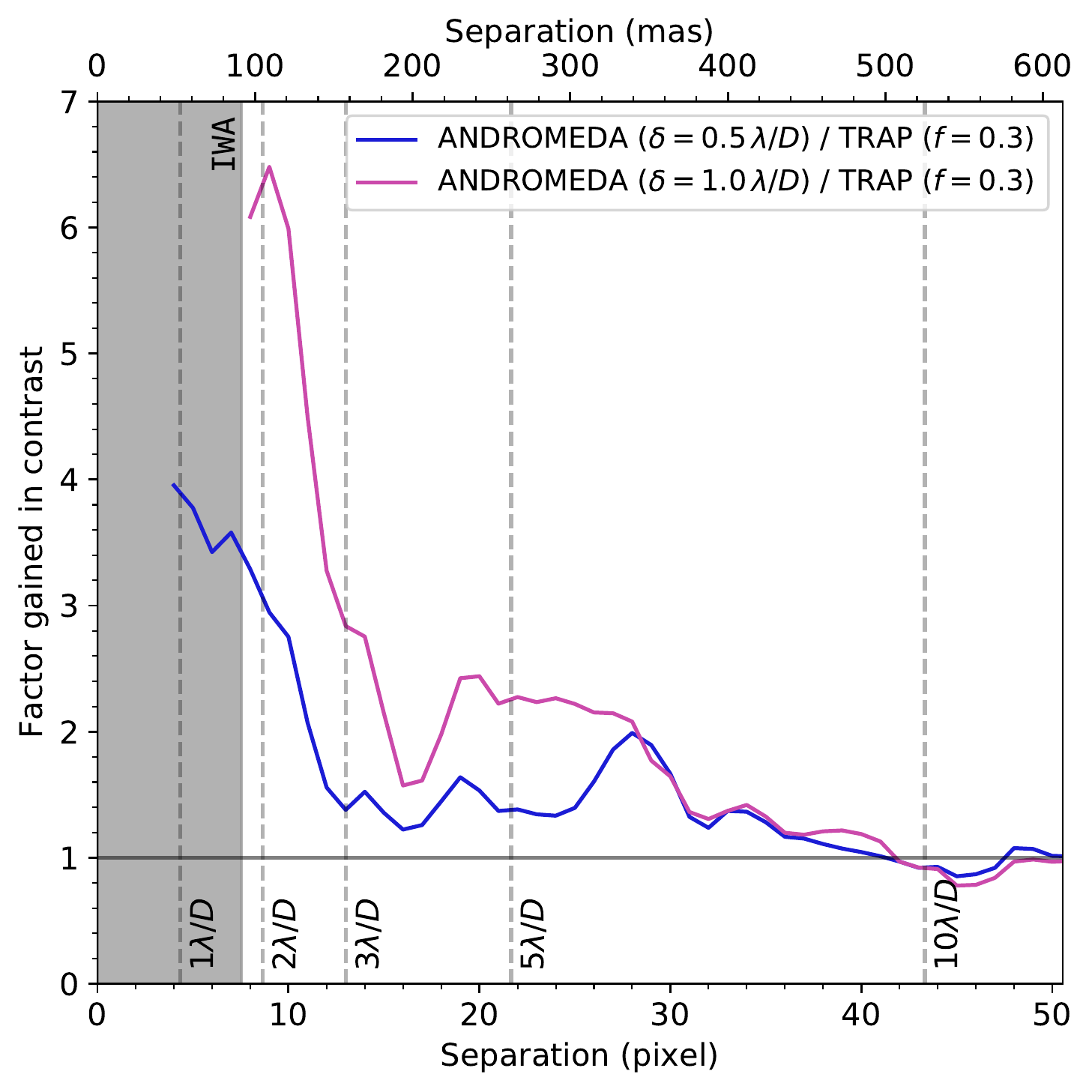}
        \end{tabular}
        \caption[Detection limit comparison between TRAP and ANDROMEDA for 51~Eri~b]{\small{Comparison between the contrast obtained with TRAP and two ANDROMEDA reductions for 51~Eri using the same input data for the K1 band. Here, TRAP was run with 30\% of available principal components, whereas the two ANDROMEDA reductions correspond to a protection angle of $\delta = 0.5\, \lambda/D$ and $\delta = 1.0\, \lambda/D$. Separations below the inner-working angle of the coronagraph are shaded and should only be interpreted relative to each other, not in terms of absolute contrast, because the impact of coronagraphic signal transmission is not included in the forward model of either pipeline. (left) The shaded areas around the lines correspond to the 16\%--84\% percentile intervals of contrast values at a given separation. (right) Factor in contrast gained using TRAP.}}
        \label{fig:contrast_51eri}
\end{figure*}

\subsubsection{Comparison to ANDROMEDA}
The detection maps for both TRAP and ANDROMEDA are shown in Fig.~\ref{fig:norm_detection_51eri}. The first thing we note is that the remaining structures in the detection map using TRAP are smoother and less correlated on spatial scales of $1\, \lambda/D$. Besides 51~Eri~b, we do not detect any other signal with >$5\, \sigma$ significance. Table~\ref{tab:51eri_phot} shows a summary of the obtained photometry for 51~Eri~b using TRAP and ANDROMEDA.

Figure~\ref{fig:contrast_51eri} shows the contrast curve for both reductions (left panel) and the factor gained in contrast by using TRAP compared to ANDROMEDA (right panel). The ANDROMEDA results have been obtained using two different values for the protection angle $\delta=0.5\, \lambda/D$ and $\delta=1.0\, \lambda/D$. Because both algorithms obtain a 2D detection map from the forward model grid of positions, not only can we determine the median detection limit at a given separation, but also the azimuthal distribution of detection contrasts. Thus, in addition to the median, we plot the 68\% range as shaded regions reflecting the variability of the detection contrast along the azimuth. This is another important figure of merit to evaluate the performance of the algorithms.

The detection limit obtained with TRAP is consistently lower than ANDROMEDA for separations up to about $10 \, \lambda/D$, at which point the results between TRAP and ANDROMEDA converge to the same background value.  At small separations, in particular, we can clearly see a gain in contrast due to the absence of a protection angle in our reduction. The ANDROMEDA curves cut off at about $1 \lambda/D$ and $2 \lambda/D$ respectively because no reference data that fulfills the exclusion criterion exists (see Fig.~\ref{fig:protection_angle}, $\sim 60^\circ$ elevation). The relative gain in contrast, with respect to the spatial model (ANDROMEDA) is significant, with ramifications for the detectability of close-in planets. It can be as high as a factor of six at $2 \lambda / D$ for a protection angle of $1 \lambda/D$ and four at a separation of $1 \lambda / D$ for a protection angle of $0.5 \lambda/D$.

The diminishing gain in contrast performance with separation and subsequent agreement with ANDROMEDA at about $10\lambda/D$, is consistent with our expectation that the limiting factor of insufficient FoV rotation for the spatial model ceases to be important at larger separation.
We note that in general the azimuthal variation of the contrast at each specific separation bin is consistently and significantly smaller for our model. This reflects the visual impression of a smoother detection map overall. This difference in azimuthal variation is especially noticeable at separations >$350$ mas, where the influence of the wind-driven halo declines (Fig.~\ref{fig:51eri_image}).

We performed detailed tests with injected point sources to rule out that TRAP biases the retrieved S/N to any significant degree, meaning that any bias is small compared to the derived statistical uncertainties. The description of these tests and their results are shown in Appendix~\ref{sec:injection_tests}.

\subsubsection{Changing the number of principal components}
\label{sec:principal_components}
\begin{figure}[t]
        \centering
        \includegraphics[width=\columnwidth]{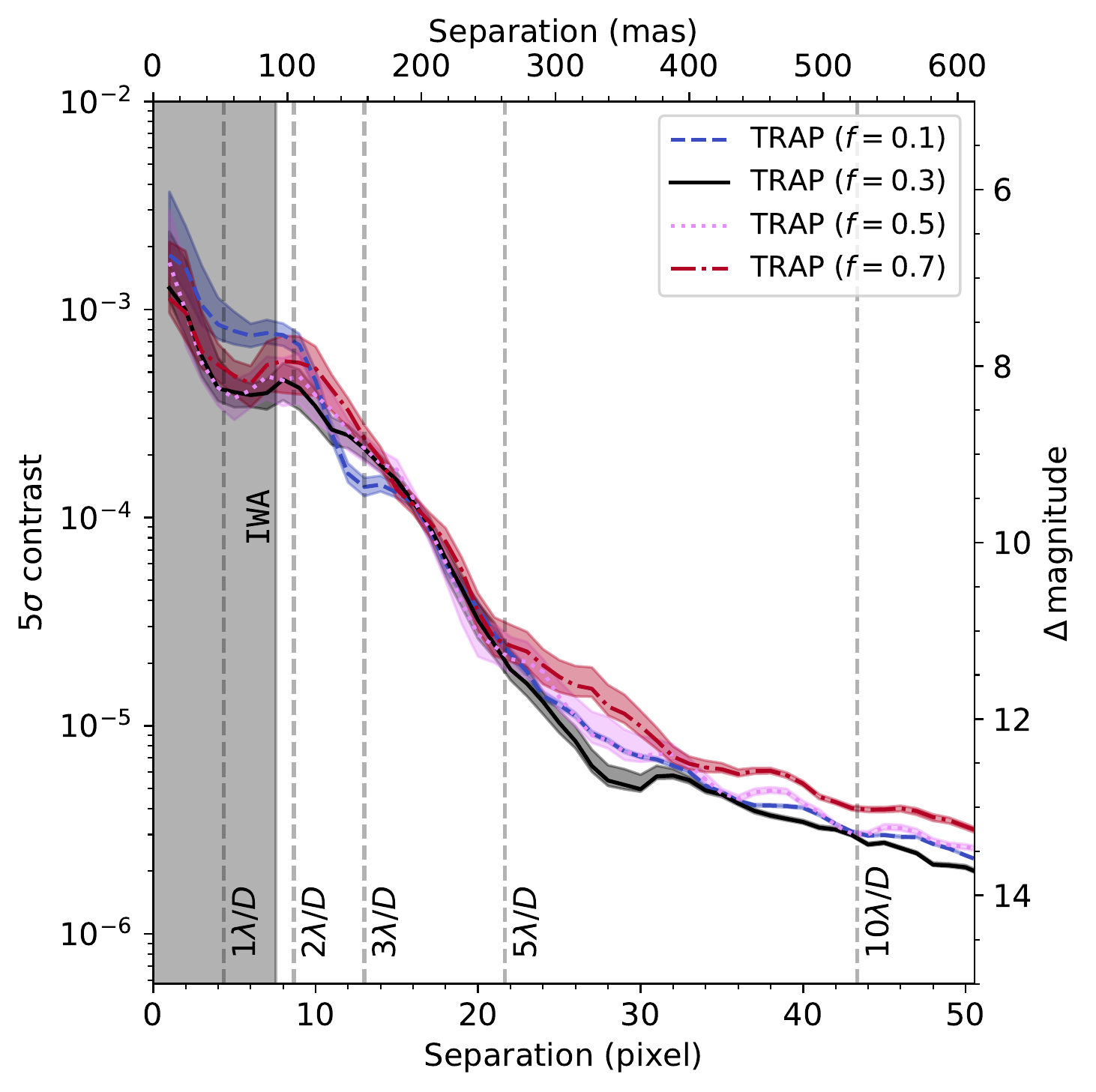}
        \caption[Detection limits for TRAP using various systematics model complexities]{Contrast curves obtained with TRAP for 51~Eri when different fractions of the maximum number of principal components are used. Figure description is analogous to that of Fig.~\ref{fig:contrast_51eri}.}
        \label{fig:fraction_dependency}
\end{figure}

To confirm the reliability of our results, we performed the same data reduction with different complexities of the systematics model. We test the impact of the fraction, $f$, of the maximum number of components, $N_\text{max} = N_\text{frames}$ (see Eq.~\ref{eq:design_matrix_2}). Figure~\ref{fig:fraction_dependency} shows the contrast curve for the same data as used in Fig.~\ref{fig:contrast_51eri}, with $f=0.1, 0.3, 0.5, 0.7$.
Firstly, we note the absence of a trend towards ``better'' detection limits with the increasing number of principal components, that is, more complex systematics models. In fact the contrast gets worse for large fractions. While a simultaneous fit of models of the planet signal and systematics counteracts overfitting, increasing the complexity of the systematics model beyond a certain point, the planet model component becomes less constraining, resulting in a larger scatter of the planet contrast prediction.
Secondly, the value of $f=0.3$ provides good results irrespective of separation, meaning that we do not have to make a significant distinction in the complexity of our temporal model depending on separation. Models that are not sufficiently complex ($f=0.1$) result in a slightly worse performance at close separation, which could be related to the presence of the strong wind-driven halo and effects from the (mis)alignment of the coronagraph with the star in addition to the quasi-static speckle pattern.

In spatial models, we may have to choose a different model complexity based on the separation to try and compensate for the self-subtraction effects (resulting from the use of smaller protection angles) by using a less complex model, but also because we have a variable number of spatial modes to reconstruct, depending on the separation and reduction area. However, for our temporal approach the existence of a model complexity that fits well globally is in agreement with our expectations. The number of frames available for training does not depend on separation because we do not have a temporal exclusion criterion. Additionally, a speckle lifetime analysis performed by \citet{Milli2016} for SPHERE did not show a strong separation dependence for speckle correlations linearly decreasing correlations on timescales of tens of minutes. The fast-evolving and exponentially decaying correlations ($\tau \sim 3.5$ seconds) that are likely to be caused by turbulence internal to the instrument did show a slight separation-dependence, but our integration times are too long and the effects too small to be expected to be visible.

\subsubsection{Changing the temporal sampling}
\begin{figure*}
        \centering
        \begin{tabular}{cc}
        \includegraphics[width=.49\textwidth]{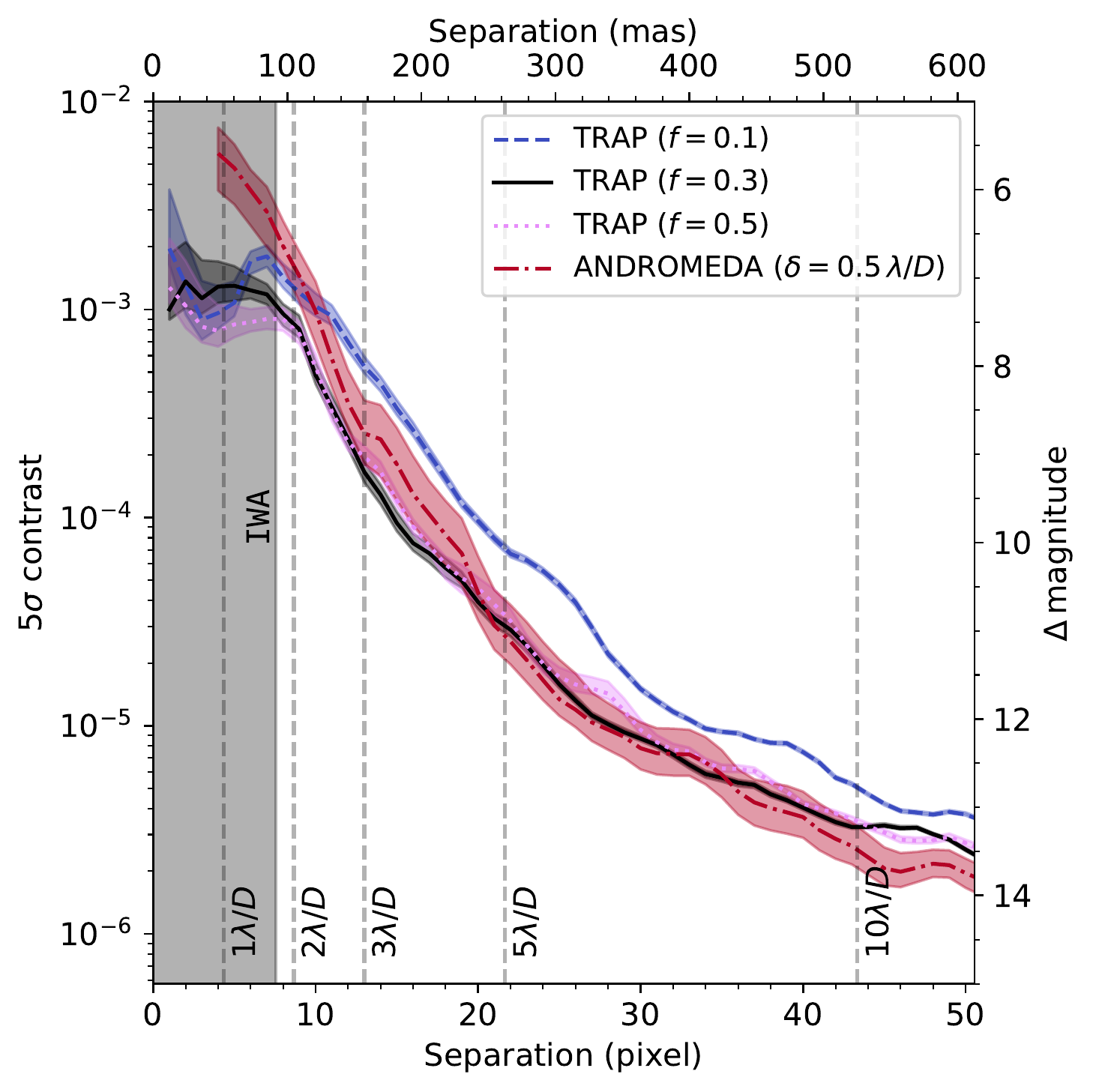}
        \includegraphics[width=.49\textwidth]{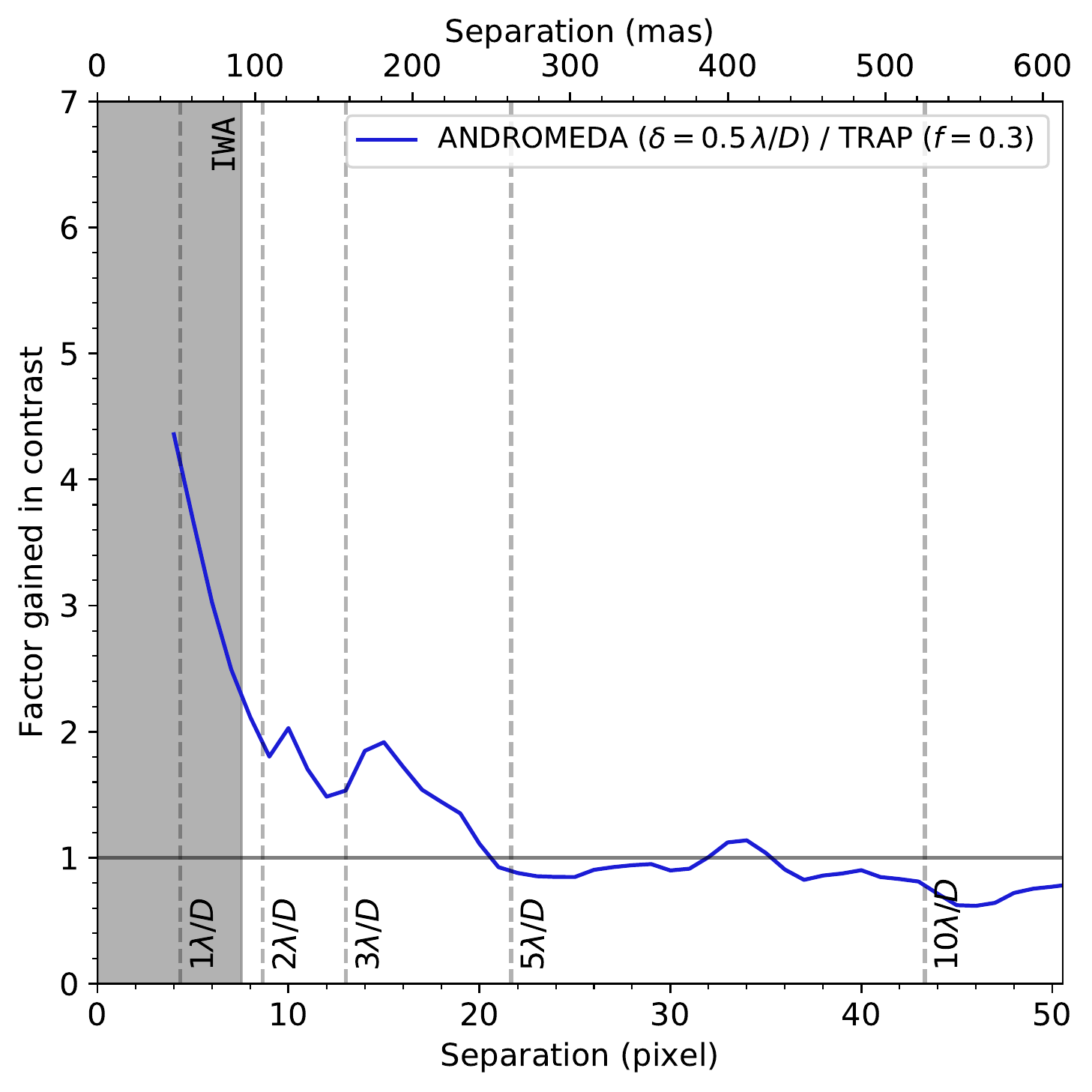}
        \end{tabular}
        \caption[Detection limit comparison between TRAP and ANDROMEDA for 51~Eri~b for temporally binned data I]{\small{Same as Figure~\ref{fig:contrast_51eri}, but each four frames of the 51~Eri~b coronagraphic data are temporally binned to achieve a DIT of 64s. Figure description is analogous to that of Fig.~\ref{fig:contrast_51eri}.}}
        \label{fig:contrast_51eri_bin4}
\end{figure*}

For a temporal systematics model, it is highly probable that the performance of the algorithm scales with temporal sampling. In Fig.~\ref{fig:contrast_51eri_bin4}, we repeat our reduction with the same dataset binned down by a factor of four ($16\,$s to $64\,$s exposures). We again plot the contrast of TRAP reductions with varying principal component fractions, $f$, and ANDROMEDA with the standard setting of $\delta = 0.5\, \lambda/D$ (left panel) for the same binned data. We also plot the factor gained in contrast compared to ANDROMEDA (right panel). We can still see an improvement at small separations, but the advantage of using a temporal model drops off faster, and it may even perform worse at large separations. From $\sim 5 \lambda / D$, we do not see any improvement and the improvement at small separations is smaller. This is consistent with our expectation of temporal models being able to capture more systematic variations at a finer time sampling. If the time sampling is poor the causal relationships in the systematics get averaged out and become more difficult to model.

\begin{figure}
        \centering
        \includegraphics[width=\columnwidth]{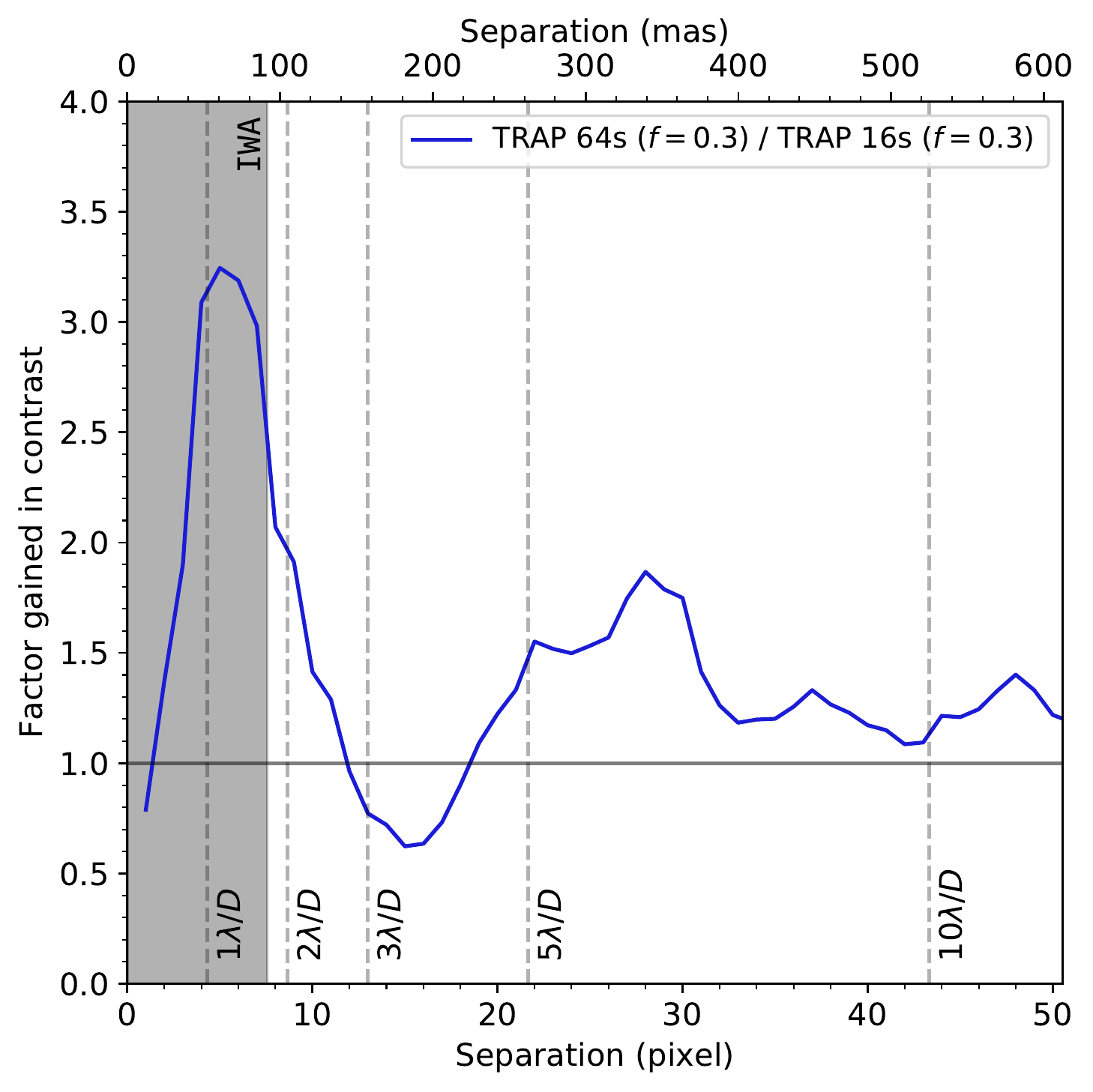}
        \caption[Detection limit comparison between TRAP and ANDROMEDA for 51~Eri~b for temporally binned data II]{Effect of using faster temporal sampling DIT 16s (unbinned) compared to slower sampling (longer averaging) with DIT 64s (binned) on contrast obtained with TRAP. Figure description is analogous to that of Fig.~\ref{fig:contrast_51eri}.}
        \label{fig:binning_effect}
\end{figure}

Figure~\ref{fig:binning_effect} shows the advantage in contrast when using the unbinned ($16\,$s exposure) compared to the binned ($64\,$s exposure) data. We see that temporally binning the data has an adverse effect on the contrast over almost the whole range of separations. Over separations between 0.5 and 10 $\lambda / D$, we see an average 40\% improvement in contrast by using the faster temporal sampling. We note that both of the sampling rates shown here are too coarse to model the short-lived speckle regime, and we expect further improvement by reducing the exposure time by another factor of four or more ($\leq4\,$s), as shown in the next section.

\begin{table}[t]
\caption[Photometry and SNR for 51~Eri~b]{Photometry and S/N for unbinned 51~Eri~b data}
\centering
\begin{tabular}{c c c}
\hline\hline
Method & Contrast & S/N \\
& ($10^{-6}$) & \\
\hline
TRAP&$6.7 \pm 0.8$ & 9\\
ANDROMEDA&$6.3 \pm 0.8$ & 8\\
\hline
\end{tabular}
\label{tab:51eri_phot}
\end{table}

\subsection{$\beta$~Pic: continuous satellite spot data with short integrations}
\begin{figure*}[t]
        \centering
        \includegraphics[width=\textwidth]{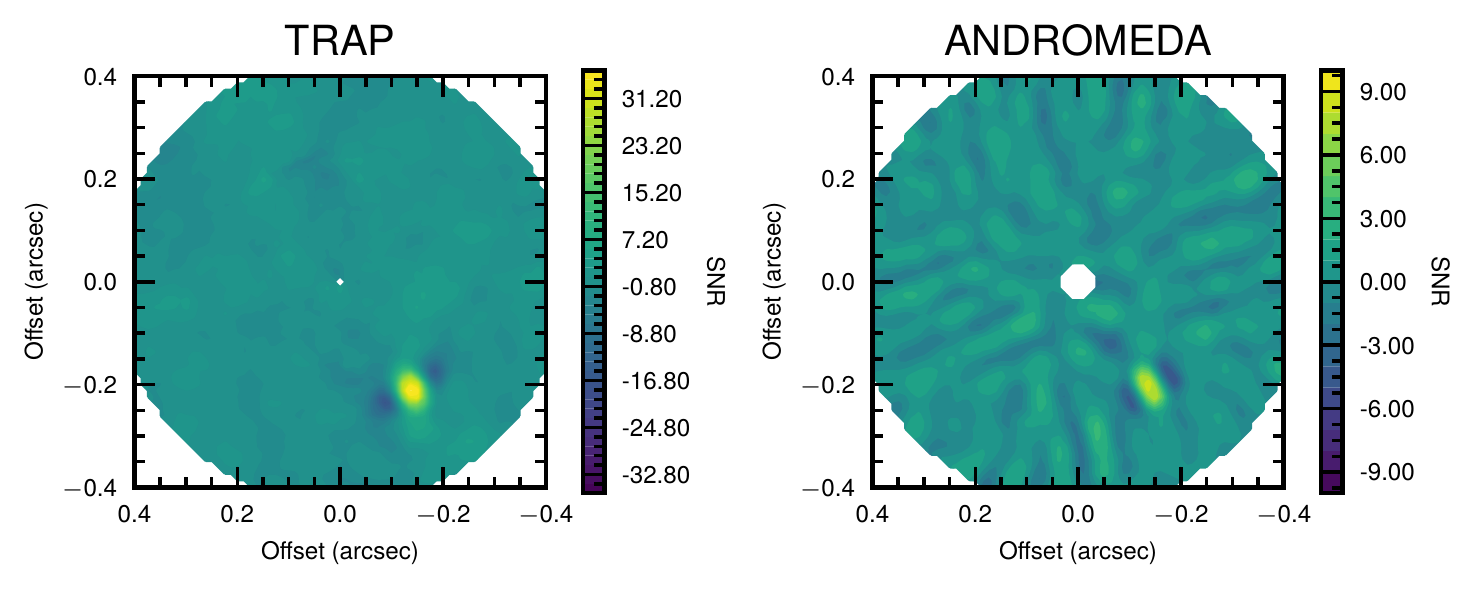}
        \caption[Detection map comparison between TRAP and ANDROMEDA for $\beta$~Pic]{Contour map of normalized detection maps obtained with TRAP ($f=0.3$) and ANDROMEDA ($\delta = 0.5\,\lambda/D$) for $\beta$ Pic. These maps must not be confused with derotated and stacked image. They represent the forward model result for a given relative planet position on the sky ($\Delta$RA, $\Delta$DEC), i.e. the conditional flux of a point-source predicted by the forward model given a relative position, corresponding to a trajectory over the detector (all pixels affected during the observation sequence).}
        \label{fig:norm_detection_betapic}
\end{figure*}

A comparison between the detection maps obtained with TRAP and ANDROMEDA is shown in Fig.~\ref{fig:norm_detection_betapic}. For both reductions, we used the same pre-reduced and aligned input data cubes and the reduction parameters were the same as for the above 51~Eri data ($f=0.3$ for TRAP and $\delta=0.5\, \lambda/D$ for ANDROMEDA). We focus our discussion on one of the two channels (H2). The color scaling differs to account for the difference in the S/N of the detection. With an S/N of 40, the TRAP detection is about four times higher than in ANDROMEDA. The S/N in H3 (not shown here) is even slightly higher because the contrast to the host star is more favorable at these wavelengths. Table~\ref{tab:betapic_phot} shows a summary of the photometry results obtained for all reductions of $\beta$~Pic~b discussed in this section.

\begin{figure*}
        \centering
        \begin{tabular}{cc}
        \includegraphics[width=.49\textwidth]{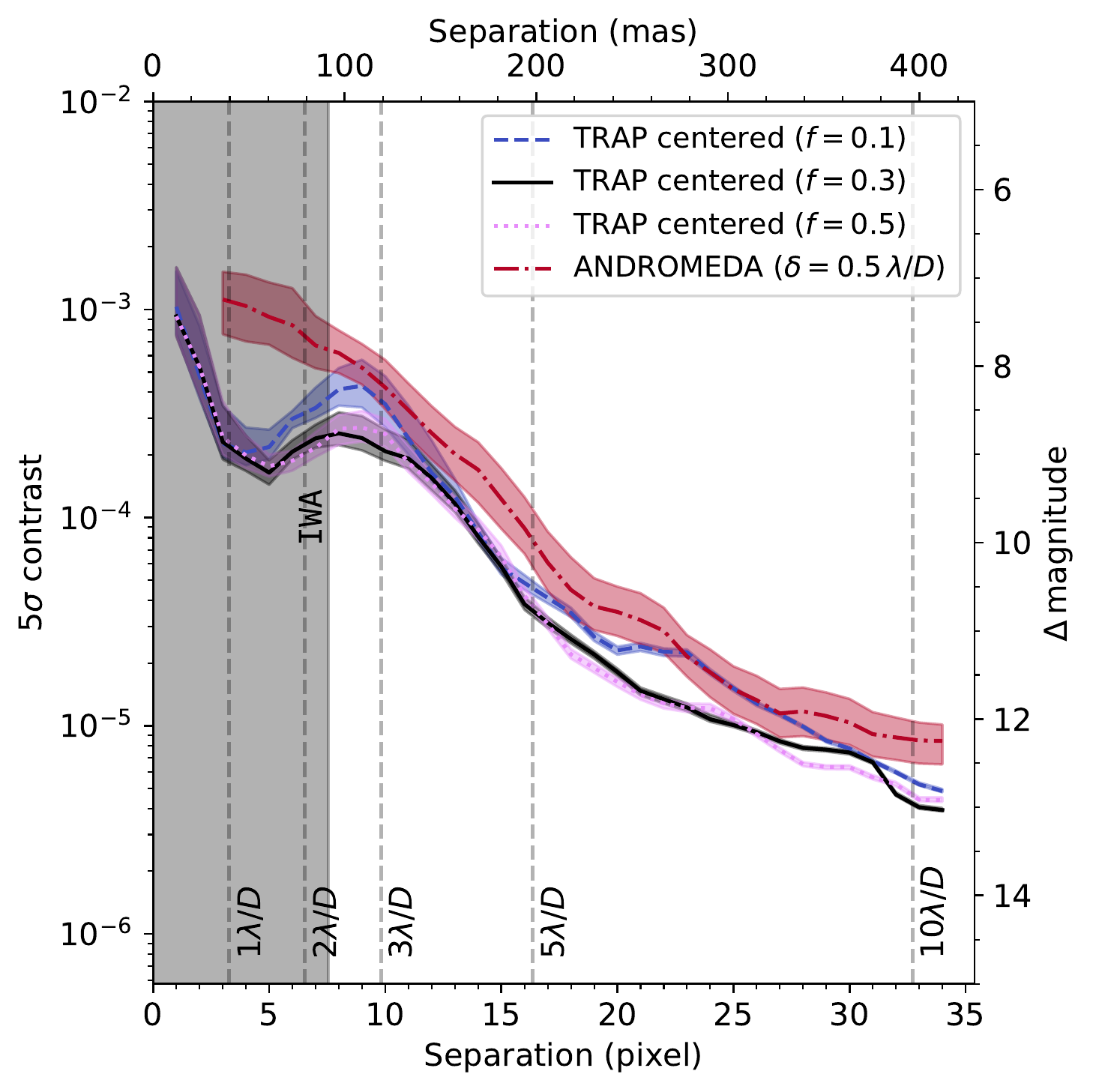}
        \includegraphics[width=.49\textwidth]{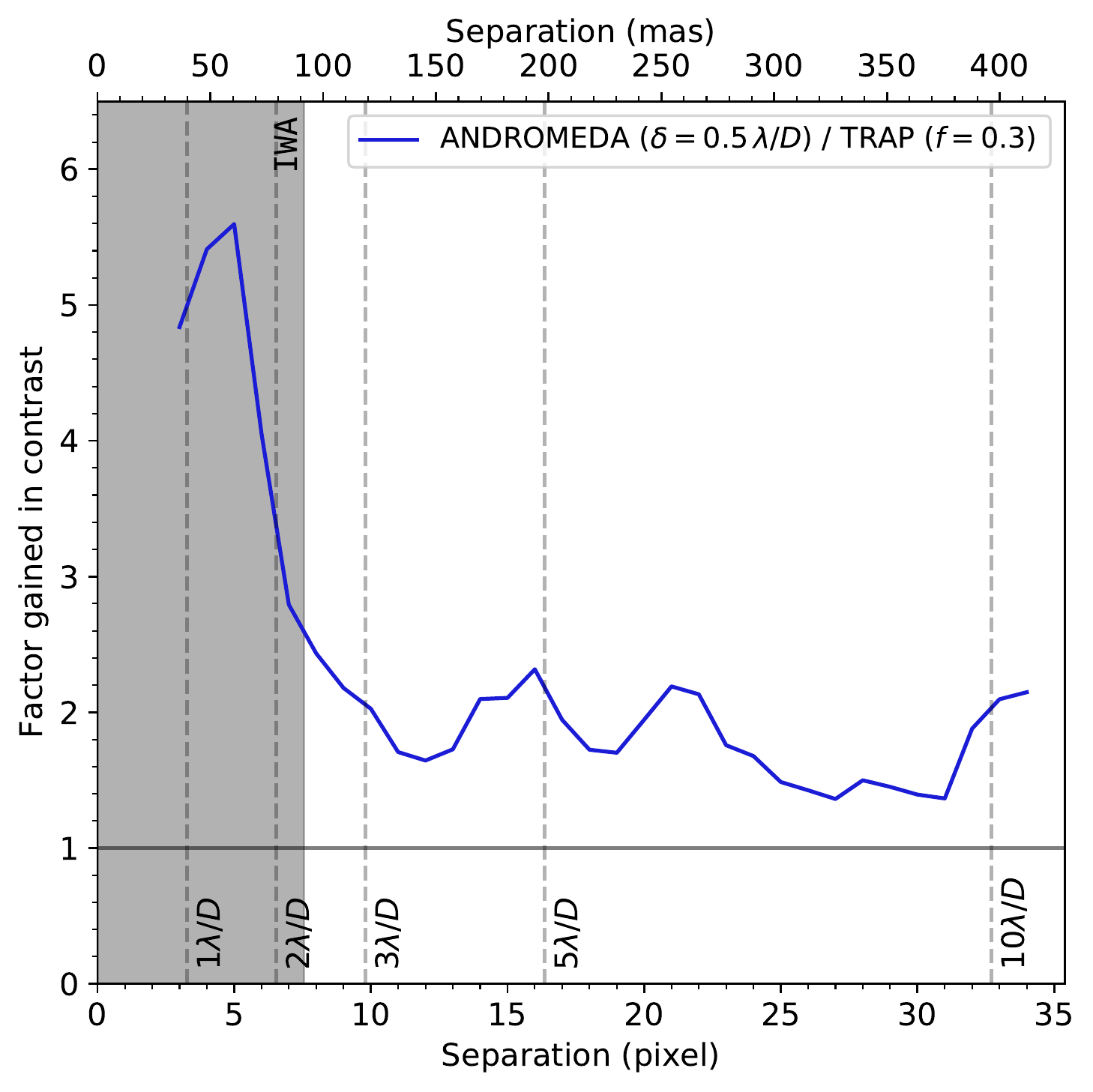}
        \end{tabular}
        \caption[Detection limit comparison between TRAP and ANDROMEDA for $\beta$~Pic]{Comparison between the contrast obtained with TRAP and ANDROMEDA reductions for $\beta$ Pic using the same input data for the H2 band. TRAP has been run with 10\%, 30\%, and 50\% of available principal components, whereas the ANDROMEDA reduction correspond to a protection angle of $\delta = 0.5\, \lambda/D$. Separations below the inner-working angle of the coronagraph are shaded and should only be interpreted relative to each other, not in terms of absolute contrast, because the impact of coronagraphic signal transmission is not included in the forward model of either pipeline. (left) The shaded areas around the lines correspond to the 14\%--84\% percentile interval of contrast values at a given separation. (right) Contrast gained by using TRAP.}
        \label{fig:contrast_betapic}
\end{figure*}

\begin{figure}[t]
        \centering
        \includegraphics[width=\columnwidth]{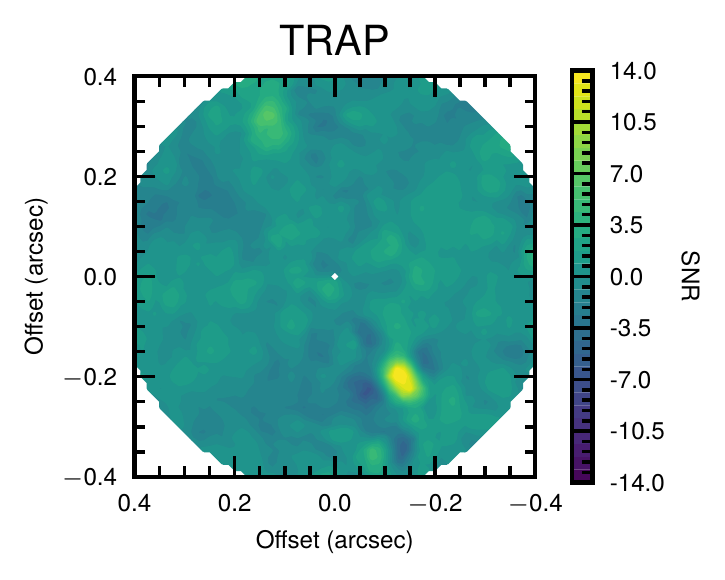}
        \caption[Detection limit comparison between TRAP and ANDROMEDA for $\beta$~Pic for temporally binned data I]{Contour map of normalized detection maps obtained with TRAP ($f=0.3$) binned data of beta Pic (16x binning, 64s exposures). Figure description is analogous to that of Fig.~\ref{fig:contrast_51eri}.}
        \label{fig:norm_detection_betapic_binned}
\end{figure}

\begin{figure}[t]
        \centering
        \includegraphics[width=\columnwidth]{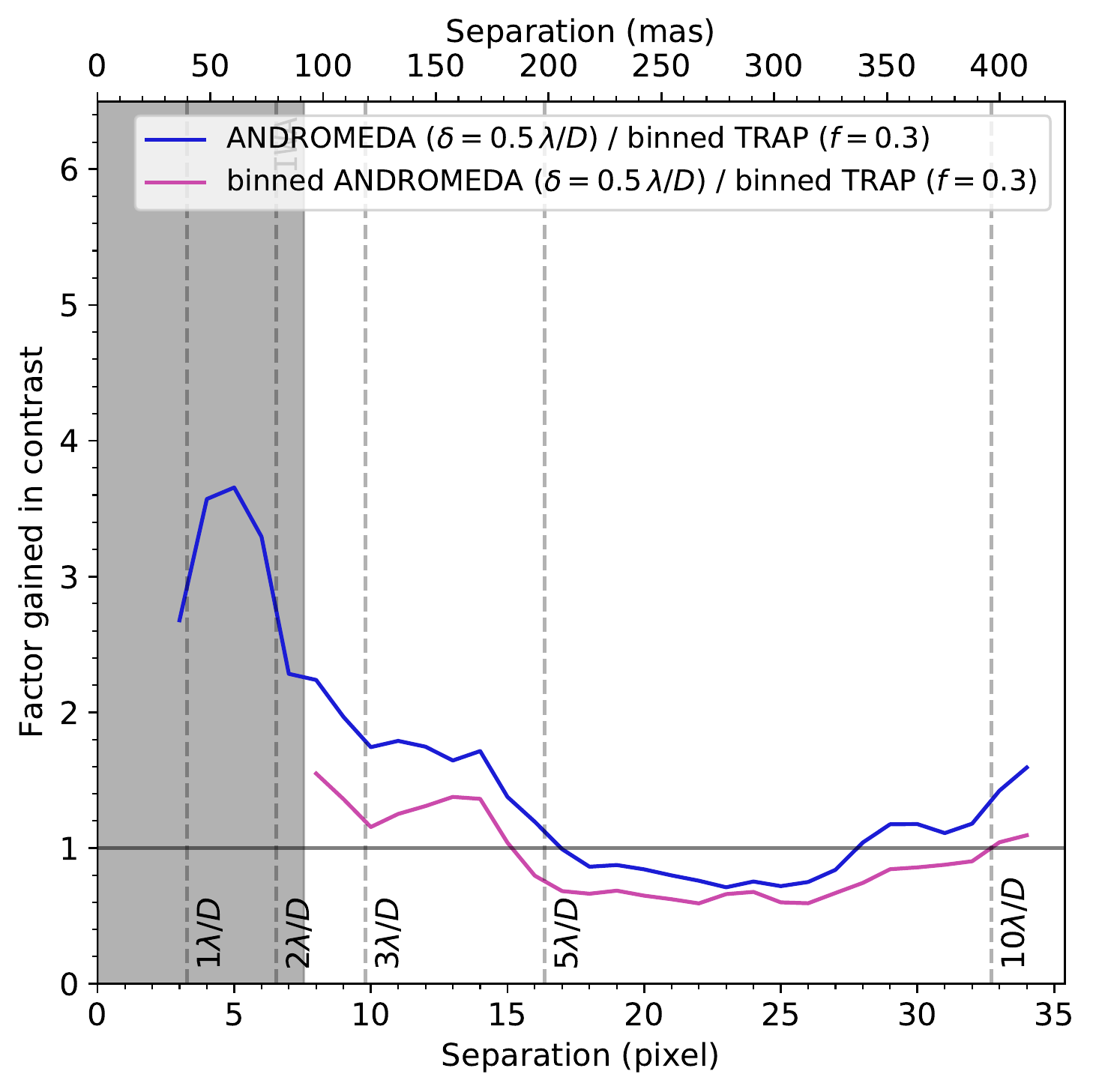}
        \caption[Detection limit comparison between TRAP and ANDROMEDA for $\beta$~Pic for temporally binned data II]{Contrast ratio between TRAP ($f=0.3$) reduction on temporally binned data (16x binning, 64s exposures) and ANDROMEDA reduction of the same binned data, as well as unbinned data, of $\beta$~Pic. Figure description is analogous to that of Fig.~\ref{fig:contrast_51eri}.}
        \label{fig:contrast_betapic_binned}
\end{figure}

Figure~\ref{fig:contrast_betapic} shows the obtained contrast analogous to Fig.~\ref{fig:contrast_51eri}. Qualitatively, we see the same effects as for 51~Eri, an increasingly significant contrast improvement at small separations from using a temporal model. We see an even more pronounced reduction in azimuthal variation in contrast, that is, the ``width'' of the contrast curve. For this data, we additionally see a baseline increase in performance between 50 -- 200\% at larger separations ($>3\,\lambda/D$) that we attribute to the better temporal sampling. To confirm this hypothesis, we have binned down the data by a factor of 16 to obtain one-minute exposures. The detection map obtained with TRAP is shown in Fig.~\ref{fig:norm_detection_betapic_binned} and the contrast compared to the unbinned ANDROMEDA reduction is shown in Fig.~\ref{fig:contrast_betapic_binned}. The planet signal is detected with a S/N of about 13, only slightly better than the S/N obtained with ANDROMEDA on the unbinned data. The obtained detection contrast curve is also comparable to ANDROMEDA at this separation. We therefore attribute the S/N improvement by a factor of four to the fact that our algorithm is capable of taking into account the bulk of short-timescale variations. We note that at the separation of $\beta$~Pic~b for the epoch of the observation ($\sim$6$\,\lambda / D$), the exclusion time for spatial models is on the order of minutes even for very small protection angles ($\delta=0.3\, \lambda/D$) and above ten minutes for the standard setting of $\delta=0.5\, \lambda/D$ (see Fig.~\ref{fig:protection_angle}), which is much longer than the exposure time.

\subsection{Applying the algorithm to unaligned data}
\begin{figure*}[t]
        \centering
        \includegraphics[width=\textwidth]{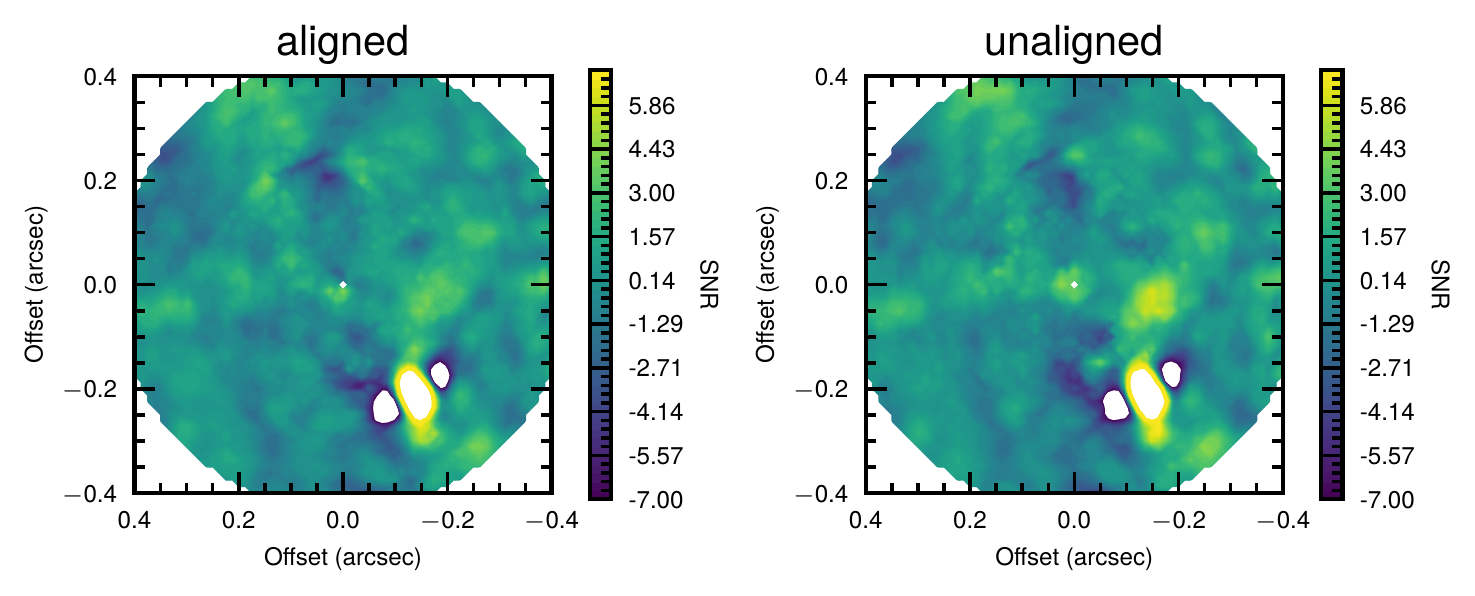}
        \caption[Detection map comparison between TRAP and ANDROMEDA for $\beta$~Pic for aligned and unaligned data]{Contour map of normalized detection maps obtained with TRAP ($f=0.3$) on aligned and unaligned data for beta Pic. These maps must not be confused with derotated and stacked image. They represent the forward model result for a given relative planet position on the sky ($\Delta$RA, $\Delta$DEC), i.e. the conditional flux of a point-source predicted by the forward model given a relative position, corresponding to a trajectory over the detector (all pixels affected during the observation sequence).}
        \label{fig:norm_detection_betapic_comparison}
\end{figure*}

As we are using a non-local, temporal systematics model and are not attempting to reconstruct a spatial model for how the speckles ``look,'' we can forego aligning the data and we can run the algorithm on minimally pre-reduced (background\footnote{The background subtraction is needed for the unsaturated PSF model. For the coronagraphic sequence this may not be true, because we already include a free constant offset term in the design matrix (see Eq.~\ref{eq:design_matrix_1}~and~\ref{eq:design_matrix_2}). Furthermore, variations of the thermal background on large scales may be incorporated in the temporal systematics model.} and flat corrected) unaligned data. To demonstrate this, we measure the center based on the satellite spots for each frame of the $\beta$~Pic data and use this varying center position to construct the light-curve model for the planet, that is, we do not shift the frames; instead, we shift the forward model for our planet. We also apply the anamorphism correction for SPHERE to the relative position of the planet by reducing the relative separation of the model by 0.6\% in y-direction \citep{Maire2016b} for each frame, instead of stretching the images. We also modulate the contrast of the planet model by the satellite spot amplitude variation measured for each frame.
The result is shown in Fig.~\ref{fig:norm_detection_betapic_comparison}, with the aligned data on the left and the unaligned data on the right. We note that the S/N of the detection is virtually the same. There are only slight differences in the residual structures. We do see a blob above the position of $\beta$~Pic~b that edges above $5\,\sigma$ in the reduction on unaligned data. It is difficult to evaluate the veracity of these structures due to the presence of the disk structure in the $\beta$~Pic system.

\begin{figure*}
        \centering
        \begin{tabular}{cc}
        \includegraphics[width=.49\textwidth]{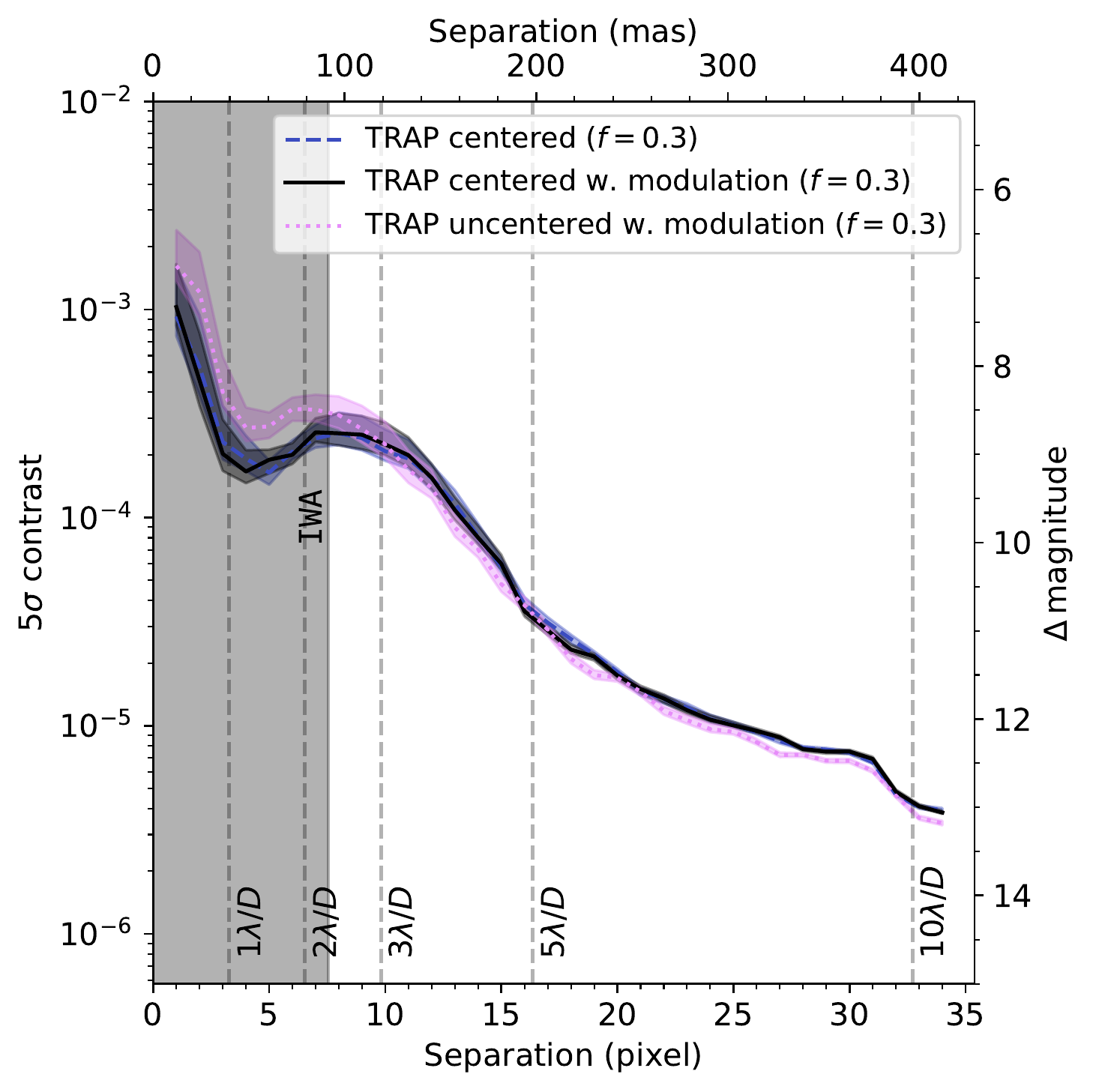}
        \includegraphics[width=.49\textwidth]{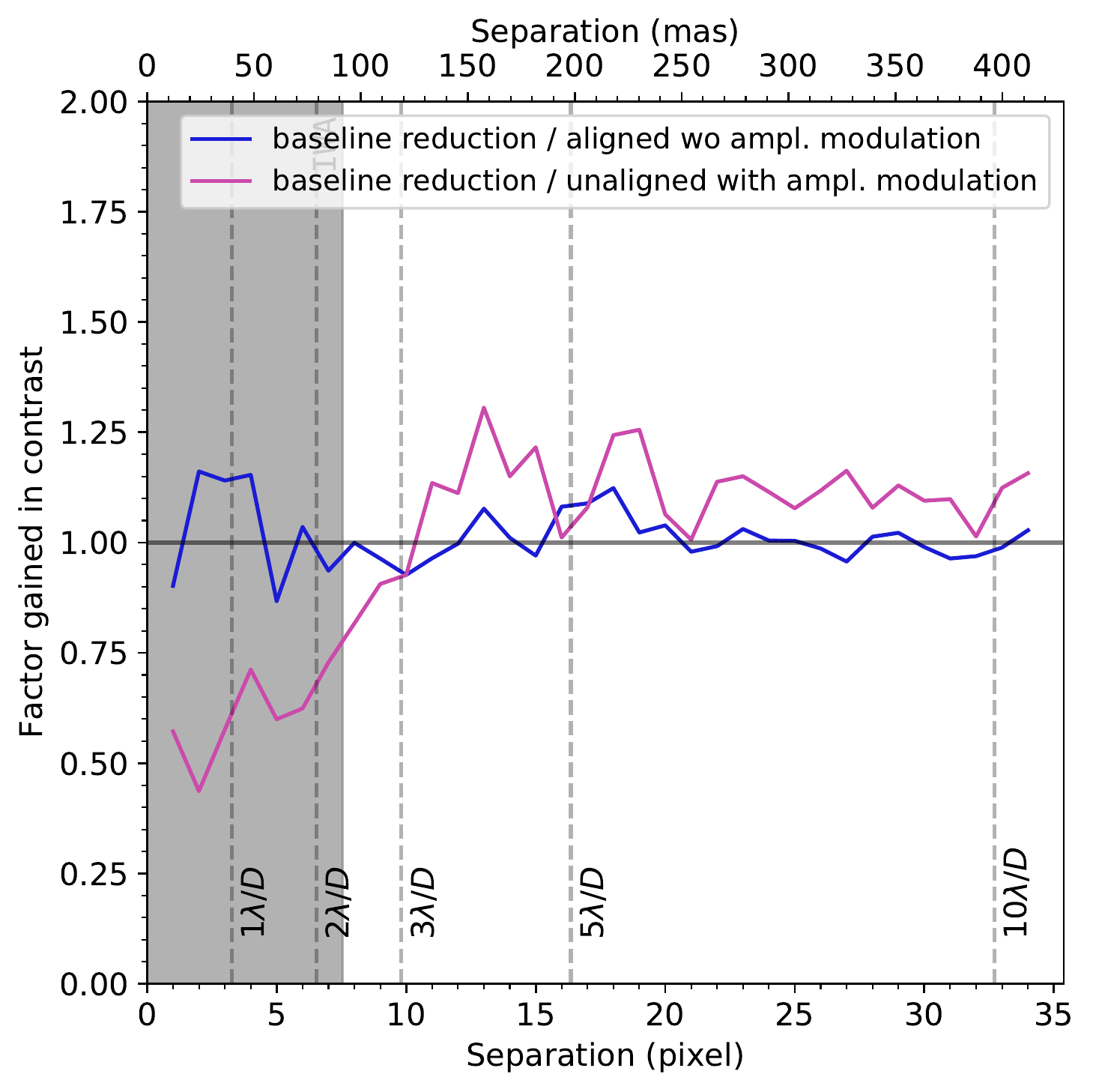}
        \end{tabular}
        \caption[Detection limit obtained with TRAP for $\beta$~Pic depending on alignment and PSF brightness modulation]{(left) Comparison between the contrast for $\beta$~Pic obtained with TRAP on: 1)~aligned data not taking into account the brightness modulation; 2)~same but taking  into account the brightness modulation; and 3)~unaligned data with the brightness modulation. TRAP was run with 30\% of available principal components. The shaded areas around the lines correspond to the 14\%-84\% percentile interval of contrast values at a given separation. (right) Contrast gain compared to aligned data without including amplitude variations (baseline reduction). Figure description is analogous to that of Fig.~\ref{fig:contrast_51eri}.}
        \label{fig:contrast_betapic_comparison}
\end{figure*}

In Fig.~\ref{fig:contrast_betapic_comparison}, we show the contrast curves for the: 1) aligned data without planet brightness modulation; 2) aligned data with planet brightness modulation; and 3) unaligned data with planet brightness modulation. The step of taking into account the brightness modulation derived from the satellite spots does not have a noticeable impact on the contrast limits. It does have a minimal impact on the derived contrast of $\beta$~Pic~b as it reduces the flux calibration bias incurred by assuming an average or median contrast for the planet flux in the planet model, instead of a more realistic distribution. In the case of this observation, the scatter of brightness variation is roughly Gaussian with a variability of $\sim$6\% centered on the mean of the satellite spot brightness, without a large systematic trend. Taking  this variation into account becomes more important as the conditions become more unstable and when an overall trend is present, for example, a trend that results from clouds reducing atmospheric transmission during part of the observation.

A bigger difference can be seen in the reduction of the unaligned data. The contrast appears to be worse in the innermost region covered by the coronagraphic mask, but slightly better outside of $3\lambda / D$. The S/N of $\beta$~Pic~b, again, is virtually the same as in the aligned case, but astrometry and photometry are slightly altered.
\begin{table}[t]
\caption[Photometry and SNR for $\beta$ Pic b]{Photometry and S/N for $\beta$ Pic b}
\centering
\begin{tabular}{c c c c c c}
\hline\hline
Method & Mod. & Align & Bin &  Contrast & S/N \\
& & & 16x & ($10^{-4}$) & \\
\hline
TRAP&no&yes&no&$1.33 \pm 0.04$ & 38\\
TRAP&yes&yes&no&$1.32 \pm 0.04$ & 38\\
TRAP&yes&no&no&$1.24 \pm 0.03$ & 38\\
TRAP&no&yes&yes&$1.29 \pm 0.09$ & 15\\
ANDROMEDA&no&yes&no&$1.19 \pm 0.13$ & 9\\
ANDROMEDA&no&yes&yes&$0.96 \pm 0.12$ &8\\
\hline
\end{tabular}
\label{tab:betapic_phot}
\tablefoot{
\footnotesize
Overview of reduction results depending on whether satellite spot amplitude modulation, pre-aligned (centered) data, and temporal binning is performed. All TRAP reductions used $f=0.3$ and ANDROMEDA reductions $\delta = 0.5 \, \lambda/D$.}
\end{table}

\subsection{Computational performance}
The computational time needed to reduce the 51~Eri dataset (256 frames) at one wavelength up to a separation of $\sim$45 pixel ($\sim$550 mas) for standard parameters ($f=0.3$, PSF stamp size 21x21 pixel), while including the variance on the data, is about 120~minutes on a single core on a laptop (intel CORE i7 vPro, 8th Gen; 16 GB memory). The computational time can be roughly halved by using a PSF stamp of half the size for determining the reduction area, $\setPy$ (excluding the first Airy ring). This only has a minor impact on the overall performance of the algorithm. Reducing $f$ similarly reduces the computational time, which is important for large data sets with thousands of frames. The algorithm is parallelized on the level of fitting the model contrast for a given position, such that the grid of positions to explore is divided among the available cores. At the current version of the code, using four cores reduces the time needed to about 40 minutes, that is, by a factor of three, due to inefficiencies in memory sharing. We expect a nearly linear relation with the number of cores after improvements to the code's parallelization architecture.

\subsubsection{Scaling with number of frames}
It is noteworthy that our algorithm's computational speed scales better with the number of frames in the observation sequence than traditional spatial-based approaches that include a protection angle. The absence of a temporal exclusion criterion means that the principal component decomposition has to be performed only once for one assumed companion position, instead of having a separate training set for each frame in the sequence. The only increase in computational time stems from the need of decomposing a larger matrix once per model position and subsequently inverting a larger system of linear equations for each pixel. Having tested the computational time for different temporal binning factors, we note that the computational time scales between linear and quadratic with the number of frames with a power-law index of about $t \propto N_\text{frames}^{1.5}$, that is, doubling the number of frames more than doubles the computational time.

\subsubsection{Scaling with the outer-working angle}
The algorithm selects a new set of reference pixels depending on the tested companion position because we have to exclude the reduction area from the training set. As such, the time spent on constructing the model is linearly proportional to the number of positions tested, which, when exploring a linear parameter space in $\Delta$RA and $\Delta$DEC, means that the number of PCAs performed is proportional to the search area, such that $N_\text{PCA} \propto r^2$, where $r$ is the separation from the central star. At the same time the number of pixels affected by a potential companion also increases with separation $N_\text{pix, affected} \propto r$. In terms of computational time, however, the time spend on PCA is relatively minor (once per tested position), and the scaling with number of affected pixels that need to be fit outweighs. Testing the algorithm with increasing outer-working angle (OWA), we derive a power-law index of about $t\propto \text{OWA}^{2.5}$.
If computation time is an issue, our algorithm can easily be used for the inner-most region exclusively and combined with the results of an algorithm that scales better with separation further out, as we do not expect substantial performance improvements at large separations with TRAP.

\section{Discussion}
\label{sec:discussion}
We have seen that a purely temporal model can be an alternative to a purely spatial model, but it is clear that neither one is a complete solution. There is a large unexplored space of spatio-temporal mixture models that would simultaneously take into account temporal and spatial correlations. 
One way of implementing such spatio-temporal hybrid models can be thought of as extending the data vectors from either time series of individual pixels or images at specific times to time series of patches of pixels. In such a model, the time series of one patch can be re-constructed in a basis set of vectors each containing the time series of another patch of equal size that are taken from a different part of the image (non-local model). Such an approach would optimize both spatial and temporal similarity between multiple such patches. It may also be possible to improve the results of TRAP by fitting a spatial LOCI-like systematics model to the image residuals after subtracting the temporal systematics model or vise-versa. Spatial and temporal regression approaches are not mutually exclusive. They can be synergistic because they optimize their models based on different correlations and different training data. This is in contrast to applying a spatial model iteratively \citep{Brandt2013}.

\subsection{Applicability to extended objects}
The topic of disks is one aspect of non-local models and regressor selection that has not been discussed in this work and requires future research. Protection from self-subtraction by using non-local training data, along with protection from over-subtraction by simultaneously fitting a forward model could be a valuable property for disk imaging, where preserving the morphology of the object is paramount. It should be pointed out that similar to the spatial approaches (with the exception of RDI), a completely azimuthally homogeneous structure will not be picked up by our algorithm in its current form because we include a constant offset term in our fit. The current detection map based on point-source forward modeling is not suited for disk imaging and would have to be adapted. It will pick up on (non-homogeneous) disk structures in the detection map but it should not be used directly to study disk morphology.

\subsection{Future improvements}
This work demonstrates the potential of the non-local, temporal systematics modeling approach. However, there are still many possible avenues for future improvements. The next step will be extending the algorithm to take into account spectral information. This can easily be achieved by adding the light curves of pixels at other wavelengths to the reference set similar to what is done in spatial models \citep[e.g.,][]{Sparks2002, Mesa2015, Ruffio2017}.
Currently, the area of the detector affected by the signal of interest is excluded from the training data. However, because the planet position stays fixed, whereas the speckle pattern scales with wavelength, we can shift the reduction area inward or outward proportional to the wavelength and add those signal-free pixels to the training set. This adds a local autoregressive component to the model that traces the temporal behavior of the speckles at the position of interest. Again, this can be achieved without interpolating the raw data or prescribing a detailed chromatic behavior other than the rough wavelength scaling needed for the reference selection.

Another important step is to improve the fidelity of the forward model, for example, by directly including the coronagraphic throughput model \citep[e.g.,][]{Guerri2011, Mawet2013} in the forward model of the signal. This is a more consistent approach than post-hoc adjusting the contrast curve by the coronagraphic transmission. As we push towards smaller inner-working angles, a good understanding of the coronagraph at all wavelengths will become important and should be modeled as well as measured on-sky as part of publicly available instrument calibrations.

There are a multitude of effects on the companion signal that, in principle, can be included in the forward model, such as the distortion of the PSF shape,  variations of the Strehl ratio, low-wind effect \citep{Cantalloube2018, Cantalloube2020}, smearing caused by integration time \citep{Lafreniere2007}, and optical aberrations as, for example, measured by a focal plane wavefront sensor \citep{Wilby2017}. Likewise, currently our algorithm implicitly assumes that all necessary information on the systematic temporal trends is encoded in other pixels. It has been shown for transit photometry that ``missing'' information on systematic trends can be accounted for using auxiliary data or a Gaussian process trained on auxiliary data \citep{Gibson2014}.
It is possible that including additional external information (e.g., on the wind, state of the AO, position of derotator, temperature, focal-plane wavefront sensing)\ could further improve the algorithmic performance significantly.

The temporal modeling approach may also prove beneficial for instruments with large pixel scales, as well as undersampled PSFs, as is sometimes the case for IFUs. Large pixels or spaxels can exacerbate the problems caused by insufficient field-of-view rotation. In general the algorithm introduced in this work should be more robust against problems associated with small FoV rotation because we do not use a temporal exclusion criterion.

An interesting use case, that has not been explored in this work, is the application to space-based observatories. For observations taken at different roll angles we can also build a temporal forward model that would take the form of a step-function for affected pixels. This, again, would allow us to take into account systematics in real-time (e.g., from instrument jitter) because we do not need to exclude any frames (roll angles) from the training data.\\

The algorithm as introduced in this work is optimized for companion searches over a grid of possible positions. If our goal is the detailed characterization of a planet of a known position -- or even a disk -- this approach may not be optimal. Future developments will include exploration of specific models in a detailed characterization step using sampling approaches such as MCMC or nested sampling to obtain photometric and astrometric information, as well as their covariance, simultaneously \citep[e.g.,][]{Wang2016b}. This would allow both the exploration of more complex physical models, such as forward models of debris disks \citep{Olofsson2016}, as well as more sophisticated systematics models.
 The use of nested sampling, combined with a negative model injection approach, would provide an easy way to include all dimensions of the data in the forward model \citep[e.g., including the spectrum of the planet][]{Ruffio2017} and perform a direct model comparison based on Bayesian evidence, as is often done in transit and radial velocity detections \citep[e.g.,][]{Espinoza2019}.

\section{Summary and conclusions}
\label{sec:paper2_conclusions}
In this work, we present a new paradigm of using a temporal, non-local systematics models to more effectively search for point sources at the most scientifically interesting small separations, where traditional ADI algorithms have issues by design. This new method allows us to address a persistent problem in high contrast imaging: the lack of good and uncontaminated training data at small separations. By building a non-local, causal model, we show that we can circumvent the problem of self-subtraction entirely, while still presenting a powerful model for the systematics that are limiting our ability to detect planets. By its nature, the time-domain model is sensitive to instantaneous changes in observing conditions, which is not the case in spatial approaches that need a reference library of images sharing the same speckle characteristics. Furthermore, the temporal model is not contingent on the data being perfectly aligned: as such interpolations and re-sampling of the data in space (and wavelength, for future additions to the method that will include this dimension) can be avoided entirely.
Our implementation, called TRAP, is open-source and publicly available.

We have shown on two datasets that the TRAP pipeline performs as well or better than a similar spatial approach with a strong improvement in contrast by a factor between 1.5 and 6 at angular separations $<3\,\lambda / D$. Beyond this separation, the improvement strongly depends on the temporal sampling of the observation sequence. The azimuthal variance of the achieved contrast across all separations is strongly reduced using the temporal systematics model compared to a spatial systematics model. Increasing the integration time from $16\,$s to $64\,$s for the 51~Eri~b dataset leads to a decrease in average contrast gained by $\sim$40\%.

For short integration times ($4\,$s, $\beta$ Pic), we can achieve a significant overall improvement of the contrast by a factor of two, even at separation between 3 -- 10 $\lambda / D$. The S/N measured for $\beta$~Pic~b significantly increased from about 10 with the spatial model to about 40 by making full use of the systematics information present in the data on the short time scales. We conclude that the effect of exposure time on the achievable contrast is underexplored in the literature. Spatial algorithms currently employed are not able to make optimal use of the information contained in short time-scale variations, because typical exclusion timescales are significantly larger than the exposure times.

Our results show that fitting the planet and systematics model simultaneously constitutes a self-regulating process on the achieved contrast when we increase model complexity: increasing the systematics model complexity (i.e., the number of principle components used) does not automatically lead to ``better'' contrasts, highlighting the benefit of a combined model fit.

We demonstrate that our temporal approach can be applied to minimally pre-reduced data without aligning the frames. This is achieved by adjusting the forward model position that generates the companion light curve according to the star's center position and the anamorphism, while excluding all bad pixels from the training and reduction sets instead of interpolating them. This reduction achieves very similar results to elaborately pre-reduced data and reduces the need for intrusive data manipulation steps (interpolation, resampling). This property can be particularly useful when taking into account data uncertainties. Another benefit is a strong reduction in the processing time needed for alignment and bad pixel interpolation, which can take significant resources for datasets with many exposures or working on entire surveys. The ability to post-process data without re-sampling could prove beneficial for SPHERE-IFS in the future, because the instrument uses a hexagonal lenslet geometry. The output images have to be re-sampled to a rectilinear grid for traditional post-processing pipelines. With TRAP, we have the capability to perform the analysis on the native image geometry.
Like ANDROMEDA, our algorithm does not require the derotation of frames. Spatial filtering, which improves ANDROMEDA performance and is also used in pyKLIP, is not needed for our algorithm to perform well.

We do not recommend dithering for pupil-tracking data as it is not necessary and can interfere with the performance of temporal models.
We further recommend the use of continuous satellite spot mode to improve the forward model performance with accurate center and amplitude variations. We strongly recommended exploring shorter integration times for observation sequences. Decreasing the integration time of IRDIS from $64\,$s to $4\,$s increases the observation overheads by about 15 percent points. If our scientific interest is focused on companions at small separations, we are in the speckle limited regime and the increase in read-noise is likely to be negligible, which is easily balanced by the many-fold increase in algorithmic performance.

Lastly, future and current development and deployment of coronagraphs with smaller inner-working angles will further increase the importance of this class of algorithms.

\begin{acknowledgements}
This work has made use of the SPHERE Data Centre, jointly operated by OSUG/IPAG (Grenoble), PYTHEAS/LAM/CeSAM (Marseille), OCA/Lagrange (Nice) and Observatoire de Paris/LESIA (Paris). 
We thank P. Delorme and E. Lagadec (SPHERE Data Centre) for their efficient help during the data reduction process. SPHERE is an instrument designed and built by a consortium consisting of IPAG (Grenoble, France), MPIA (Heidelberg, Germany), LAM (Marseille, France), LESIA (Paris, France), Laboratoire Lagrange (Nice, France), INAF–Osservatorio di Padova (Italy), Observatoire astronomique de l’Universit\'e de Gen\`eve (Switzerland), ETH Zurich (Switzerland), NOVA (Netherlands), ONERA (France) and ASTRON (Netherlands) in collaboration with ESO. SPHERE was funded by ESO, with additional contributions from CNRS (France), MPIA (Germany), INAF (Italy), FINES (Switzerland) and NOVA (Netherlands). SPHERE also received funding from the European Commission Sixth and Seventh Framework Programmes as part of the Optical Infrared Coordination Network for Astronomy (OPTICON) under grant number RII3-Ct-2004-001566 for FP6 (2004–2008), grant number 226604 for FP7 (2009–2012) and grant number 312430 for FP7 (2013–2016). T.H. acknowledges support from the European Research Council under the Horizon 2020 Framework Program via the ERC Advanced Grant Origins 83 24 28. We thank F. Cantalloube, and B. Pope for their comments. J.B. acknowledges support from the European Research Council under the European Union's Horizon 2020 research and innovation program ExoplANETS-A under grant agreement No.~776403.
\end{acknowledgements}


    \bibliographystyle{aa} 
    \bibliography{trap} 

\clearpage

\begin{appendix}
\section{Impact of temporal exclusion criteria on traditional ADI}
\label{sec:temporal_exclusion_impact}
\begin{figure*}[t]
  \centering
  \includegraphics[width=\textwidth]{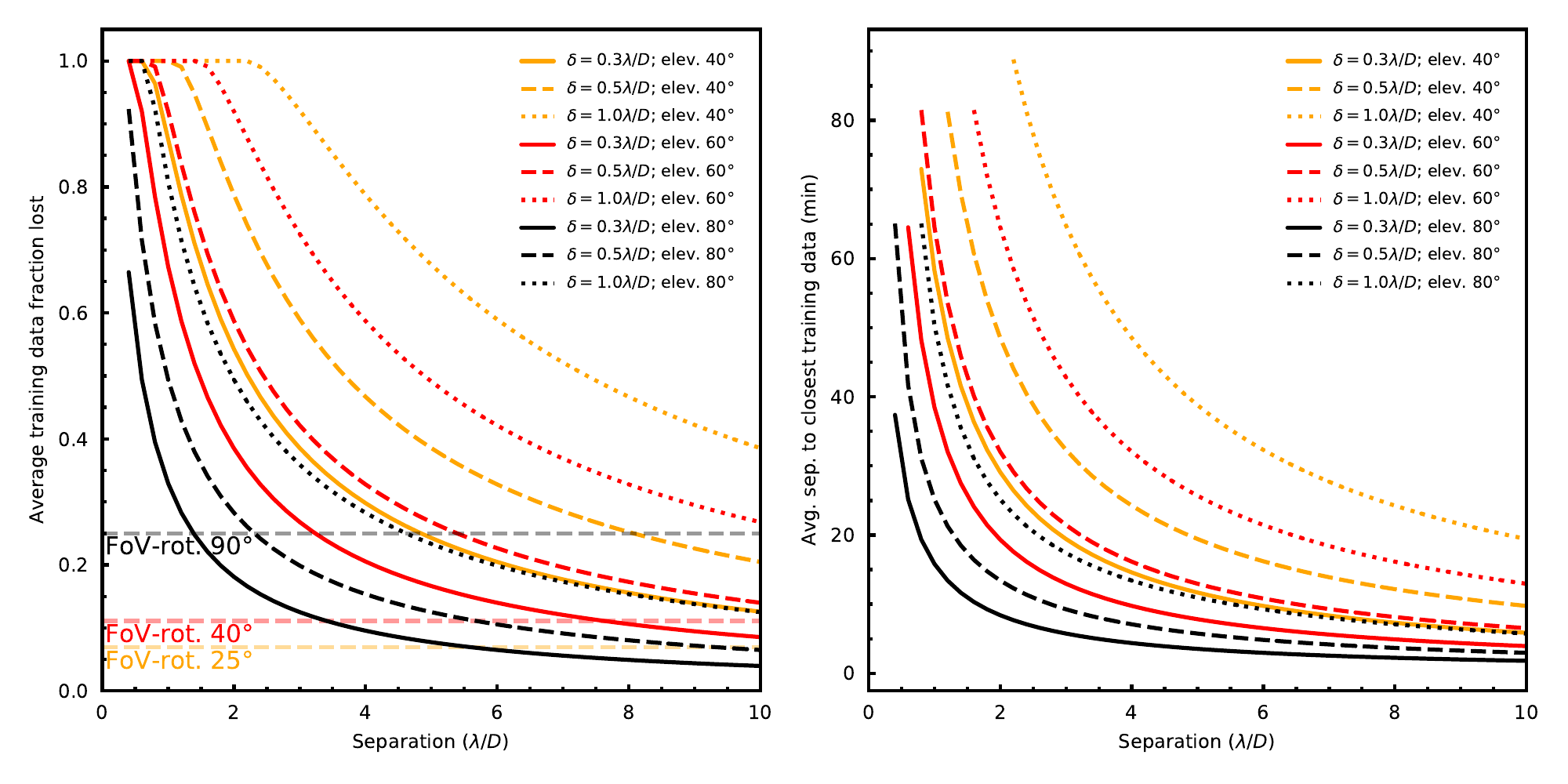}
  \caption[Temporal exclusion criterion]{Effect of a different protection angle (0.3, 0.5, 1.0 $\lambda/D$) over separation from the central star. All values are computed for an observation sequence of 90 minutes. The color of the line encodes different elevations of the target over the horizon at meridian passage. The line style gives the protection angles. The left panel shows the total fraction of training data lost due to the protection angle averaged over the whole sequence. The right panel shows the average temporal separation to the nearest viable frame outside the protection zone. It should be noted that for a temporal systematics model (horizontal lines) that excludes all pixels affected by a companion signal, the exclusion is solely determined by the field-of-view rotation. At $40^\circ$ total rotation, we would therefore exclude $\sim$10\% of the data, regardless of separation, presenting a big advantage at short separations over temporal exclusion criteria.}
  \label{fig:protection_angle}
\end{figure*}

The temporal exclusion criterion (protection angle) has a strong impact on the available data for training the systematics model in traditional spatial ADI approaches. Figure~\ref{fig:protection_angle} shows the impact of different assumed protection angles for a 90 minute observation sequence centered around meridian passage computed for targets at different maximum elevations above the horizon. The colors denotes the maximum elevation of the target ($40^\circ$, $60^\circ$, $80^\circ$), which, respectively, result in different overall FoV rotation ($25^\circ$, $40^\circ$, $90^\circ$) and correspondingly slower or faster parallactic angle change rates. The left panel shows the average training data fraction excluded due to a the chosen protection angle ($0.3\,\lambda/D$, $0.5\,\lambda/D$, $1.0\,\lambda/D$) averaged over the observation sequence. The horizontal lines corresponding to the fraction of training data lost if instead a simple spatial exclusion criterion is used. In the case of a simple spatial exclusion criterion, as used in this work, which excludes all pixels at a given separation affected by a companion signal, and assumes only pixels at a comparable separation are used to build the systematics model, the excluded training data fraction is independent of the separation and simply given by the fraction of total field-of-view rotation angle compared to $360^\circ$, that is, about 10\% for $40^\circ$ rotation. In most real-world cases, the fraction of available training data excluded by the spatial exclusion criterion is significantly less at small separations compared to a temporal exclusion criterion (protection angle).

Analogously, the right panel of Fig.~\ref{fig:protection_angle} shows the average distance in time to the closest available frame in the training set that remains available at a given protection angle. The observation sequences of targets at different maximum elevations and assuming different protection angles are shown. It is noteworthy to point out that even at larger separations this time difference is still significantly larger than typical integration times.

\section{Impact of protection angle on ANDROMEDA reductions}
\label{sec:andromeda_protection_angle}

\begin{figure*}
        \centering
        \begin{tabular}{cc}
        \includegraphics[width=.5\textwidth]{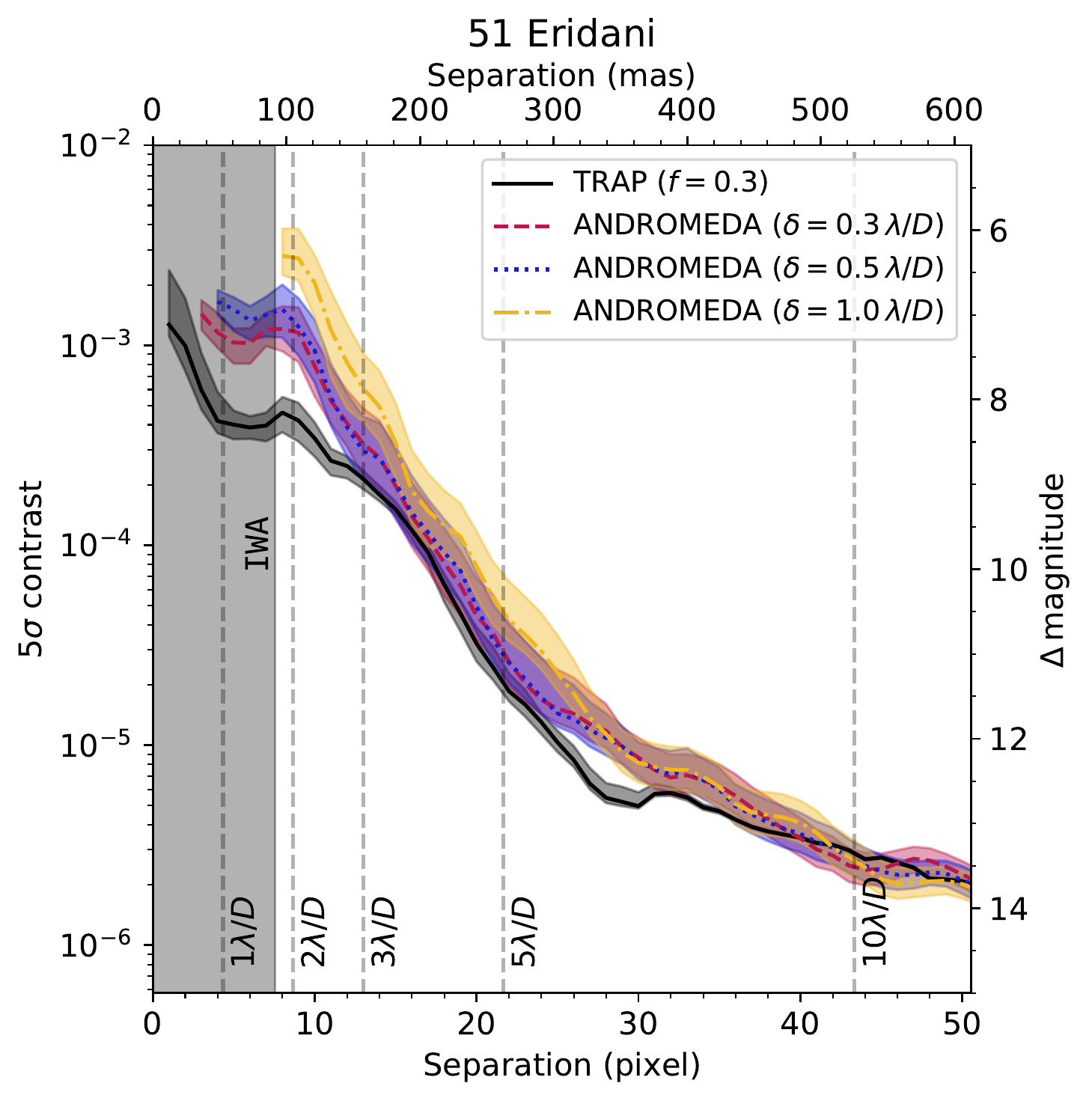}
        \includegraphics[width=.5\textwidth]{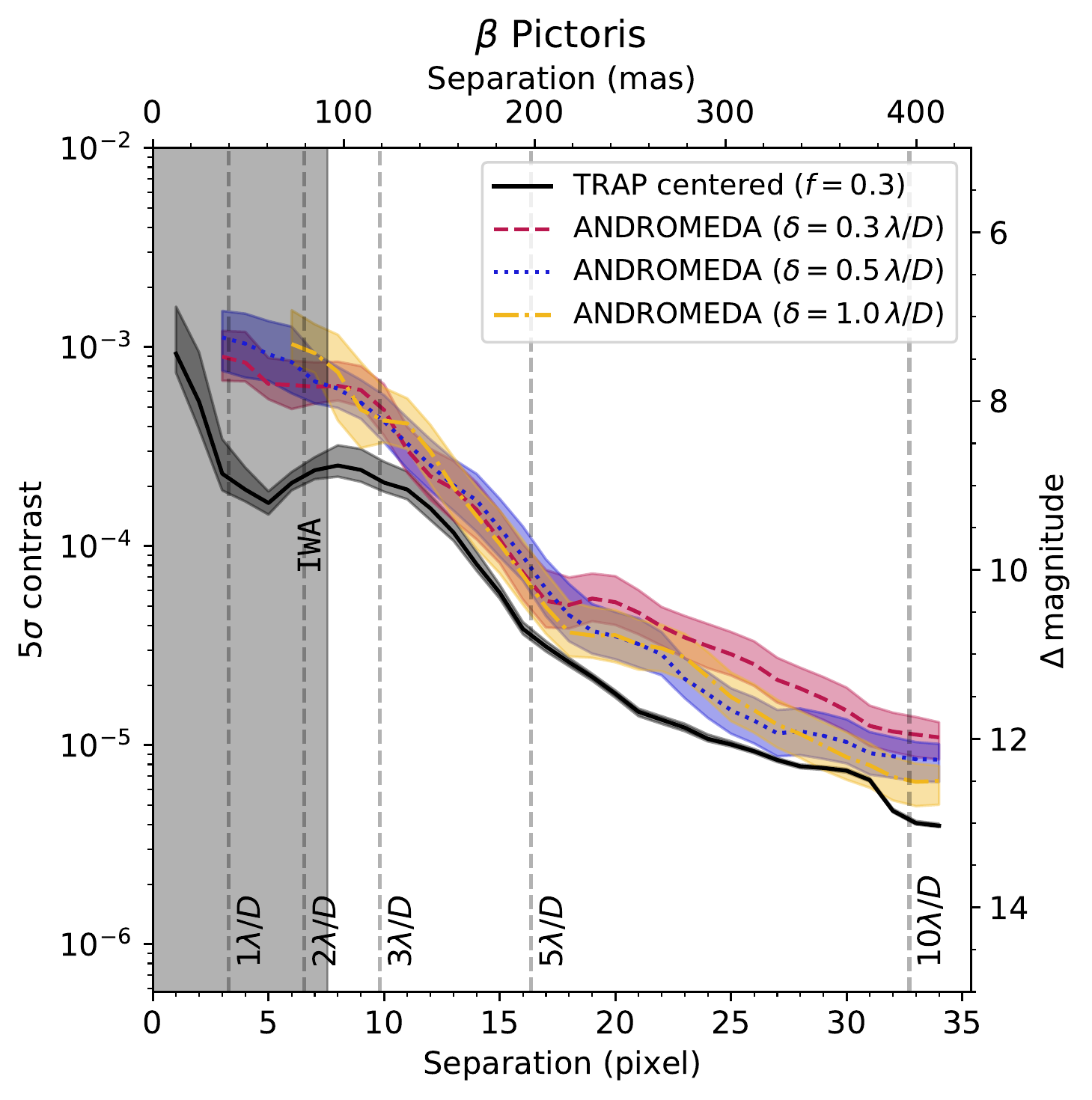}
        \end{tabular}
        \caption[]{\small{Comparison between the contrast obtained with TRAP and three ANDROMEDA reductions for 51~Eri (left) and $\beta$~Pictoris (right). TRAP has been run with 30\% of available principal components, whereas the three ANDROMEDA reductions correspond to a protection angle of $\delta = 0.3, \, 0.5, \, 1.0\, \lambda/D$. Separations below the inner-working angle of the coronagraph are shaded and should only be interpreted relative to each other, not in terms of absolute contrast, because the impact of coronagraphic signal transmission is not included in the forward model of either pipeline. The shaded areas around the lines correspond to the 16\%--84\% percentile intervals of contrast values at a given separation.}}
        \label{fig:delta_dependence}
\end{figure*}

The impact of the protection angle $\delta$ on the contrast limits obtained with ANDROMEDA is shown in Fig.~\ref{fig:delta_dependence}. In our study, we selected $\delta = 0.5 \, \lambda/D$ as being representative of the algorithm's performance, the same value as used in \citet{Samland2017}. It produces consistent and reliable results for SPHERE data at all separations. Choosing smaller angles can marginally improve performance at very small separations, but may suffer from a potential increase in systematic bias. ANDROMEDA is based on forward modeling the expected signal in difference images. For small displacements, most of the planet signal is subtracted and therefore adds more noise than signal to the analysis for faint companions. This effect is worse for data with short integration times. Large protection angles negatively impact the performance at small separations due to excluding significant fractions of the data, as discussed in length this work. All ANDROMEDA reductions use the standard spatial frequency filtering fraction of 1/4 \citep{Cantalloube2015}.

\section{Recovery of injected signals}
\label{sec:injection_tests}
\begin{figure*}
        \centering
        \begin{tabular}{cc}
        \includegraphics[width=.5\textwidth]{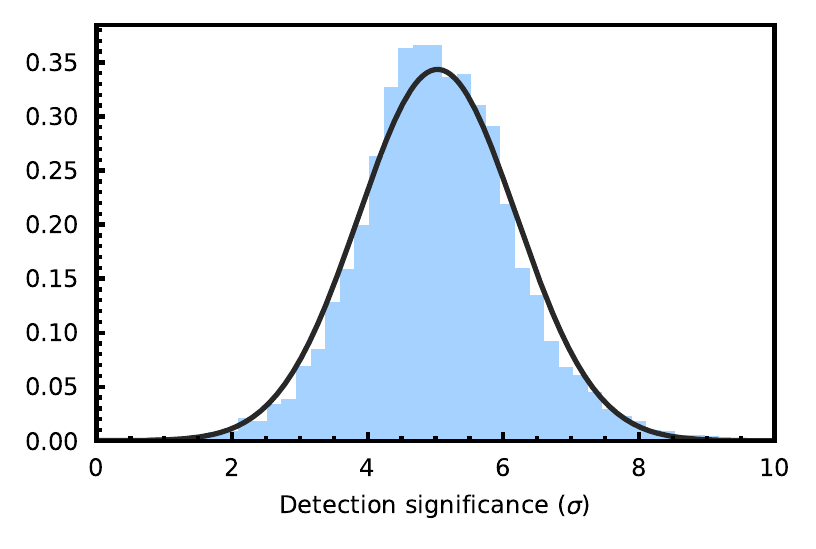}
        \includegraphics[width=.5\textwidth]{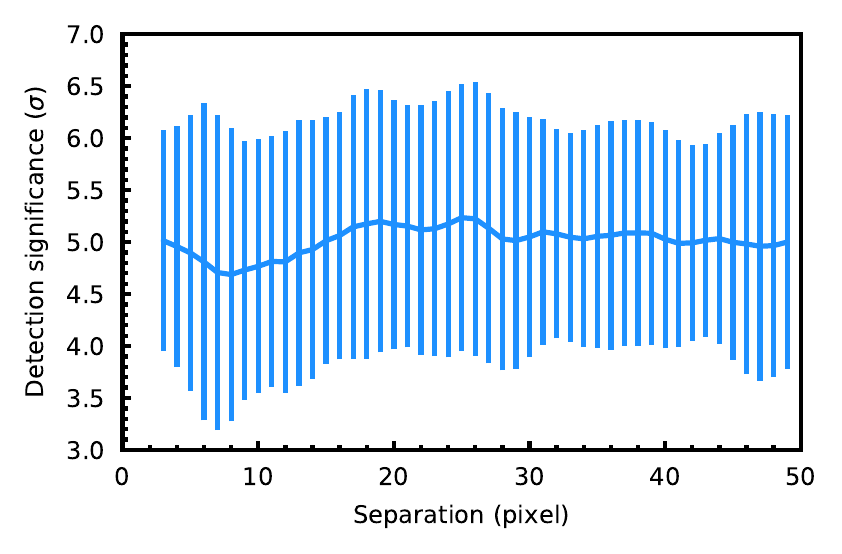}
        \end{tabular}
    \label{fig:retrieved_snr}
        \caption[]{(left) Histogram of retrieved S/N values for injected $5\,\sigma$ signals in the reduced field-of-view, overplotted with a Gaussian fit. (right) Mean and standard deviation of detection significance of injected $5\,\sigma$ signal over the separation computed in three-pixel-wide annuli.}
\end{figure*}

A common way to test the fidelity of algorithms and post-processing pipelines is to inject a known signal into the data and attempt to retrieve it. In this work, because we build a forward model of the expected set of planet light curves for each tested companion position anyway, it is a simple additional step to inject a companion model at a desired contrast into the raw data before fitting.
Additionally, as we run the algorithm for each position individually, we do not have to worry about contamination of the training set from multiple injected signals. The time necessary for performing this test is therefore only insignificantly slower than when running TRAP normally. The detailed procedure is described below:
\begin{enumerate}
    \item Create a 2D contrast and detection map following the algorithm as described in this paper. This gives the detection limit at each tested planet position.
    \item Run the algorithm again, but inject a signal for the tested position with the desired significance as determined from the contrast map. Obtain a map of retrieved contrasts for all positions.
    \item Normalize the resulting uncertainty and S/N maps with the normalization values as determined from the case without injected signal.
\end{enumerate}
This results in maps of the retrieved signal and the associated uncertainty at each position in the detection image. We performed this test on the IRDIS K2-band 51~Eri~b dataset described in this paper. The K2-band was chosen because 51~Eri~b is not detected at this wavelength, reducing potential effects of the real signal on the injected signals. The test was performed using a signal corresponding to $5\,\sigma$ for each position from a separation of four pixels ($\sim 1 \, \lambda/D$) out to a separation of 50 pixels (613~mas). The left panel of Figure~\ref{fig:retrieved_snr} shows a normalized histogram of the retrieved S/N\ over all positions, together with a fitted Gaussian distribution (mean $\mu=5.03$ and standard deviation $\sigma=1.16$). The average S/N is very close to the injected $5\,\sigma$, showing that there is weak to no systematic bias.

\begin{figure*}
        \centering
        \begin{tabular}{cc}
        \includegraphics[width=.5\textwidth]{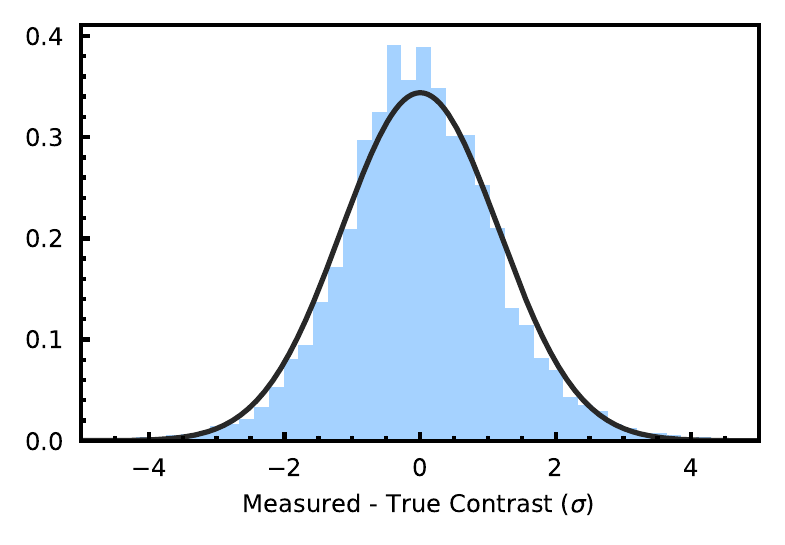}
        \includegraphics[width=.5\textwidth]{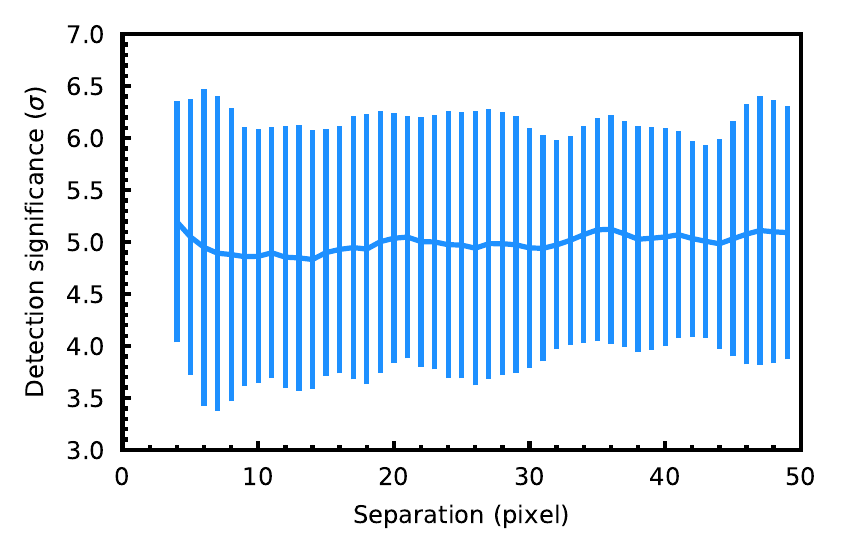}
        \end{tabular}
        \label{fig:biascorr_retrieved_contrast_deviation}
        \caption[]{(left) Histogram of deviation of retrieved contrast from true values after separation dependent bias correction for injected $5\,\sigma$ signals in the whole reduced FoV, overplotted with a Gaussian fit. (right) Mean and standard deviation of detection significance of injected $5\,\sigma$ signal over the separation computed in three-pixel-wide annuli after separation dependent bias correction.}
\end{figure*}

The right panel of Figure~\ref{fig:retrieved_snr} shows the mean and standard deviation of the distribution of retrieved S/N over separation in three-pixel-wide annuli that are analogous to the annuli used to normalize the S/N images. The mean of the detection significance stays close to the expected $5\,\sigma$ without significant systematic deviations, except for a small underestimation of the signal strength at around $2\, \lambda/D$, which corresponds to the edge of the coronagraph. Retrieved contrasts are on average the true contrasts and scatter within about $1\,\sigma$ of the values, confirming the reliability of the obtained photometric values. The remaining small biases can be understood and corrected for using the bias map, that is, the map showing the difference between the measured and injected contrast. The mean deviation from the true value per separation can be used as a bias correction of the reduction procedure and eliminate any remaining systematic overfitting or underfitting of the signal. This does not, of course, correct for biases that come from assuming an incomplete or wrong companion model in the forward model. The left panel of Fig.~\ref{fig:biascorr_retrieved_contrast_deviation} shows the recovered minus true contrast corrected by the median deviation from the true injected values of the separation bin. The mean of this distribution is $\mu = 0.002$. The scatter is unaffected. The right panel of Fig.~\ref{fig:biascorr_retrieved_contrast_deviation} shows the bias corrected detection significance over separation.

\end{appendix}
\end{document}